\documentclass[twocolumn,english,showpacs,floatfix,amsmath,amssymb,prb,superscriptaddress]{revtex4-1}

\usepackage{xcolor}
\usepackage{graphicx}% Include figure files
\usepackage{dcolumn}% Align table columns on decimal point
\usepackage{bm}% bold math
\usepackage[utf8]{inputenc}
\usepackage[T1]{fontenc}
\usepackage{mathptmx}

\begin{document}

%\preprint{AIP/123-QED}

%\title{Second-order reputation with memory in the spatial donation game.}
%\title{Effect of memory and second-order reputation on cooperation:\\a theoretical model based on experimental findings.}
\title{Effect of memory, intolerance and second-order reputation\\ on cooperation.}
% Force line breaks with \\

\author{Chengyi Xia}
\affiliation{Tianjin Key Laboratory of Intelligence Computing and Novel Software Technology, Tianjin University of Technology, Tianjin 300384, P.R.China}
\affiliation{Key Laboratory of Computer Vision and System (Ministry of Education), Tianjin University of Technology, Tianjin 300384, P.R. China}
\author{Carlos Gracia-L\'azaro}%
 \email{cgracia@bifi.es}
\affiliation{Instituto de Biocomputaci\'on y F\'isica de Sistemas Complejos (BIFI), Universidad de Zaragoza, 50018 Zaragoza, Spain}
\affiliation{Departamento de F\'isica Te\'orica, Facultad de Ciencias, Universidad de Zaragoza, 50009 Zaragoza, Spain}

\author{Yamir Moreno}
 \affiliation{Instituto de Biocomputaci\'on y F\'isica de Sistemas Complejos (BIFI), Universidad de Zaragoza, 50018 Zaragoza, Spain}
\affiliation{Departamento de F\'isica Te\'orica, Facultad de Ciencias, Universidad de Zaragoza, 50009 Zaragoza, Spain}
\affiliation{Complex Networks and Systems Lagrange Lab, Institute for Scientific Interchange, Turin, Italy}

\date{\today}% It is always \today, today,

\begin{abstract}

The understanding of cooperative behavior in social systems has been the subject of intense research over the past decades. In this regard, the theoretical models used to explain cooperation in human societies have been complemented with a growing interest in experimental studies to validate the proposed mechanisms. In this work, we rely on previous experimental findings to build a theoretical model 
based on two cooperation driving mechanisms: second-order reputation and memory.
Specifically, taking the Donation Game as a starting point, the agents
are distributed among three strategies, namely Unconditional Cooperators, Unconditional Defectors, and Discriminators, where the latter follow a second-order assessment rule: Shunning, Stern Judging, Image Scoring, or Simple Standing. A discriminator will cooperate if the evaluation of the recipient's last actions contained in his memory is above a threshold of (in)tolerance. 
In addition to the dynamics inherent to the game, another imitation dynamics, involving much longer times (generations), is introduced.
The model is approached through a mean-field approximation that predicts the macroscopic behavior observed in Monte Carlo simulations. We found that, while in most second-order assessment rules, intolerance hinders cooperation, it has the opposite (positive) effect under the Simple Standing rule. Furthermore, we show that, when considering memory, the Stern Judging rule shows the lowest values of cooperation, while stricter rules show higher cooperation levels. %May be (Carlos): These findings will help to understand the role of reputation and memory in cooperative behavior.
%OR (Chengyi): Current findings will be conducive to deeply develop the potential mechanisms to foster the evolution of cooperation.
\end{abstract}

\maketitle

%\begin{quotation}
%The ``lead paragraph'' is encapsulated with the \LaTeX\ 
%\verb+quotation+ environment and is formatted as a single paragraph before the first section heading. 
%(The \verb+quotation+ environment reverts to its usual meaning after the first sectioning command.) 
%Note that numbered references are allowed in the lead paragraph.
%
%The lead paragraph will only be found in an article being prepared for the journal \textit{Chaos}.
%Cooperative behavior in human societies constitutes a puzzle that has been both theoretically and experimentally studied in depth over the last decades.With the aim of bridging the gap between both approaches, we propose a theoretical model based on experimental findings. To this end, some agents are assumed to be hawk-eyed: when making the decision whether help or not to a given recipient, they consider not only the recipient's past actions but to whom those actions were directed. Furthermore, we consider a memory effect by assuming that reputation is based on a number $M$ of recent actions. Agents decide whether to cooperate or not according to an intolerance threshold on reputation. It is shown that both evaluation rules and intolerance %threshold have a nontrivial role in the persistence of cooperation.\end{quotation}

\section{Introduction}
The presence of cooperative behavior among unrelated individuals remains an open question in the scientific community, %been selected as one of the $25$ key scientific problems to be solved in the $21$st century 
constituting one of the current key scientific challenges \cite{pennisi2005}. The Evolutionary Game Theory \cite{axelrod06,sigmund00} provides a powerful framework to study cooperative behavior \cite{nowak_06book}, including  cooperation in structured populations \cite{szab1997,perc2010,perc2013,zhen2015_epjb}. Several mechanisms have been proposed to explain cooperation \cite{nowak2006}, such as kin selection \cite{Hamilton64}, direct \cite{Trivers71} or indirect reciprocity \cite{nowak1998a}, group selection \cite{Traulsen06}, and network reciprocity \cite{nowak1992,santos2005,yamir2007_prl,yamir2012_pnas}. Among them, indirect reciprocity does not require repeated interactions between the same pair of partners and offers a clear explanation of how this preference for cooperation has evolved \cite{nowak05_review,marsh18_review}. In a population, when an individual exhibits an altruistic behavior towards another one, he pays a cost --including time, energy or risks-- for his helping action even if he cannot get immediate returns. However, if a third party knows of his kind deed, he may provide help to this altruist in a later action so that the original cost of the first agent can be counteracted to obtain the positive benefit. That is, the helper receives the benefit not from the beneficiary himself but another individual. Indirect reciprocity requires public information about individual actions, as well as an evaluation system, so that cooperation can be sustained for a long time
\cite{ohtsuki2004,wang2012_plos,xia2016_pla}. %Without these information, it is difficult to judge what kind of action (Cooperating, $C$ or Defecting, $D$) is good or bad, and who tends to be cooperative or defective.
Thus, it is significant to build a feasible and reliable evaluation system \cite{uchida2010} to differentiate between altruistic and selfish persons, and give the corresponding reward \cite{reward1,reward2} for the contributor or punishment \cite{punish1,punish2,punish3} to the cheater.

Regarding the individual actions evaluation, probably the most popular approach is the Image Scoring, proposed by Nowak and Sigmund to explore the role of indirect reciprocity in the evolution of cooperation through computer simulations and theoretical analyses \cite{nowak1998a,nowak1998b}. They showed that cooperation can thrive via the indirect reciprocity if each agent holds an image score, being the score increased (\textit{resp}., decreased) by one point for each act of helping (not helping). According to this approach, a donor will provide help to a recipient if, and only if, this recipient has a positive score. Therefore, a player will obtain the help from others in the future if he has helped more often than he has refused to do it.

However, there is no unanimity on the effectiveness of the Image Scoring rule. As an example, Leimar and Hammerstein \cite{leimar2001} indicated theoretically that Sugden's Standing Strategy \cite{sugden1986} provided a much more effective mechanism to foster cooperation through indirect reciprocity under a more complex population structure. In Sugden's Standing Model \cite{sugden1986},
%every player starts with good standing, decreasing it only when he refuses to help a recipient with a good standing.
a player's score only decreases when he refuses to help a recipient with a good score. 
Unlike Image Scoring, defecting against a \textit{bad guy} does not penalize the donor's reputation. After that, Panchanathan and Boyd \cite{body2003} also explored the evolution of indirect reciprocity when errors are considered, showing that, under these circumstances, Image Scoring is not an evolutionary stable strategy (ESS), while the Standing Strategy can be. Henceforth, only considering the actor's action (usually termed as the first-order evaluation) is not always enough when we design the rule of reputation evaluation, 
it is necessary to take both the donor's action and the recipient's reputation into account, which is referred to as a second-order assessment rule.

As a further step, Ohtsuki and Iwasa \cite{ohtsuki2004,ohtsuki2006}exhaustively discussed the aforementioned two rules, together with other second-order reputation evaluation schemes, and found that Standing Strategy is often more successful than Image Scoring. In particular, they further pointed out that only eight cases, called ``leading eight'', significantly facilitate indirect reciprocity. At the same time, extensive experiments \cite{milinski2001,bolton2004} are also conducted to illustrate how the Standing or Scoring mechanisms are adopted in the human cooperation, and it is indicated that the Standing rule is not superior to the Scoring mechanism due to the imperfect information \cite{uchida2010_pre} or gossip \cite{Nakamaru2004} dissemination during the real experiments.

A non-negligible fact is that, with some exceptions \cite{leimar2001}, most theoretical works assume the well-mixed structure to study the reputation evaluation. Recently, Sasaki \textit{et al.} \cite{sasaki17_games} investigated the evolution of reputation-based cooperation in a regular lattice considering four leading second-order assessment rules (these rules will be defined and discussed in section \ref{Subsection:Second-orderAssesmentRules}). Through an agent-based model, they showed that those four rules lead to distinct cooperative behaviors, which strongly depends on the setup, and it is particularly indicated that the Simple Standing strategy is the most efficient one in terms of the promotion of cooperation on regular ring networks. %They also pointed out that the so-called Stern Judging rule can only foster the cooperation under the indirect and public observation for any agent's reputation.

It is worth noting that the above-mentioned theoretical models carry out the second-order reputation assessment just according to the last action of a donor and the standing status of a recipient, that is, the historical information on individual actions in those studies reduces to one step. Nevertheless, the historical information (\textit{i.e.}, memory effect) may play a role in decision-making. For instance, Wang \textit{et al.} \cite{memory_pre} presented a memory-based Snowdrift Game on top of regular lattices and scale-free networks, where the fraction of cooperating actions stored in the memory is used to determine the strategy adoption at the next generation, finding that the memory length of individuals plays
a distinct role as the cost-to-benefit ratio is changed.
%but this work does not consider the role of reputation in the cooperation.
In a recent work, Cuesta \textit{et al.} \cite{cuest15_sp} showed through experiments with memory effect that reputation fosters cooperation and drives network formation. Furthermore, they found that people measure reputation based on all the information available (memory length), giving more weight to the last action.
%Yet, they can only consider that the individual reputation evaluation is just based on the Image Scoring, that is, first-order evaluation, meanwhile the theoretical framework to model the network formation is also absent.
Thus, it is essential to combine the second-order assessment information with the memory effect to further study the role of indirect reciprocity in the evolution of cooperation, and we try to fill this gap in the current work.

The rest of this paper is organized as follows. First, in Section \ref{section:model}, we introduce the Donation Game Model with Memory and Second-Order Assessment. %, and then in detail present the iterating procedures of our model.
Secondly, in section \ref{section:MeanField}, we address the model through a mean-field
approximation that, while omitting some key features, helps its understanding and qualitatively predicts the macroscopic behavior observed in section  \ref{Section:Simulations}, where we provide the results of large numerical simulations. Finally, in section \ref{Section:DicussionAndConclusions} we discuss the implications of 
the model, together with the conclusions.

\section{Donation Game Model with Memory and Second-Order Assessment}
\label{section:model}

In this paper, we investigate the Evolutionary Donation Game in a finite size population, as formulated in most spatial indirect reciprocity models \cite{nowak05_review}. During each interaction, every actor (individual) has just one chance to play as a donor, \textit{i.e.}, donate or not to a recipient, which is chosen within his neighborhood. %Multiple interactions are used to collect the game payoff within one iterating step so that his strategy can be updated through the pairwise payoff comparison with one of nearest neighbors. Meanwhile, the individual action list 
Furthermore, individual actions history will be considered as a basis of reputation assessment \cite{cuest15_sp}, according to four typical second-order strategies \cite{ohtsuki2004,ohtsuki2006} below described in section \ref{Subsection:Second-orderAssesmentRules}. Hence, the memory effect or history of the recent actions will be combined with the reputation-based assessment rule to analyze the evolution of cooperation within a structured population. In what follows, we will describe in detail the newly proposed Donation Game Model with Memory and Second-Order Assessment.

\subsection{Donation Game}

In the proposed model, the interaction between any pair of players can be described as a Donation Game, that is, one player is selected as a donor and the other one as a recipient, and subsequently the donor will decide whether he will make a donation to the recipient or not. If he donates, the donor will pay a cost $c$, and the recipient will obtain a benefit $b$ ($b>c$)% to highlight the reciprocity since the net benefit $b-c>0$ for two players)
; if not, the donor will pay nothing, and the recipient will not receive any benefit. Although the donation does not give any direct benefit to donors, some individuals may choose to donate to show a good image and then increase their chances to get help from others in the future. Thus, the Donation Game is often chosen as a basic framework to explore the role of indirect reciprocity.

\subsection{Second-order Assessment Rules}
\label{Subsection:Second-orderAssesmentRules}
How to judge the goodness of a recipient is crucial for the here proposed model, and we select four typical second-order assessment rules as the basis of the calculation of the recipient score \cite{ohtsuki2006}. In Tab. \ref{tab1}, we depict the assessment results under these four rules including Shunning, Stern Judging, Image Scoring and Simple Standing. We summarize their main features as follows:
\begin{itemize}
  \item \textbf{Shunning}: The donor is positively evaluated when he cooperates (donates) against a cooperator. Otherwise (when he cooperates against a defector or whenever he defects), he will be negatively evaluated. This is the strictest rule to obtain a good image.
  \item \textbf{Stern Judging}: The donor will be positively evaluated if he cooperates against a cooperator, or if he defects (rejects the donation) against a defector. Otherwise, he will be negatively evaluated. To a certain extent, this rule will justify the defection since rejecting the donation to a recipient with a bad image is not considered a bad action, which helps a \textit{bad recipient} to cleanse his image by refusing to help another player with a bad image.% Stern Judging is one of the eight leading rules \cite{ohtsuki2006}.
  \item \textbf{Image Scoring}: The donor will be positively evaluated if he cooperates or negatively evaluated if he defects, regardless of the recipient's past actions. In essence, this is a first-order rule since the image of an agent is uniquely determined by his own action.
  \item \textbf{Simple Standing}: The donor will be negatively evaluated only if defects against a cooperator. Otherwise (when he defects against a defector or whenever he cooperates), he will be positively evaluated. Henceforth, the Simple Standing rule is the most tolerant rule for a donor to get a good evaluation among the four rules considered here.
\end{itemize}

For all these rules, when facing cooperator, a player will be positively evaluated if he cooperates, and negatively if he defects. The differences between these four rules appear when the donor meets a defector.%, which finally determines the evolutionary outcome. For the sake of simplicity, we also replace $G/B$ in the action list (i.e., memory) of a player with $C/D$ during the iterated procedures as described in the previous section.

\begin{table}[htbp]
\begin{center}
\begin{ruledtabular}
%\begin{tabular}{|c|c|c|c|c|}
\begin{tabular}{ccccc}
%\hline
%\multirow{2}{*}{} & \multicolumn{4}{c|}{/Actor's Strategy}       \\ \cline{2-5}
%Recipient's Image & \textbf{G} & \textbf{G} & \textbf{B} & \textbf{B}\\\hline
Recipient's Image & \textbf{C} & \textbf{C} & \textbf{D} & \textbf{D}\\\hline
Donor's Strategy      & \textbf{C} & \textbf{D} & \textbf{C} & \textbf{D} \\ \hline
Shunning          & G              & B               & B                & B              \\
Stern Judging     & G              & B               & B                & G              \\
Image Scoring     & G              & B               & G                & B              \\
Simple Standing   & G              & B               & G                & G              \\
\end{tabular}
\caption{Representative second-order reputation assessment rules. In the second row, C and D (\textit{i.e.}, cooperation and defection) designate the action of the donor facing a recipient whose previous action is displayed in the first row. From third to sixth rows, G/B denotes that the donor will be evaluated as good (G) or bad (B) after the corresponding actions.}\label{tab1}
\end{ruledtabular}
\end{center}
\end{table}

\subsection{Initial Conditions}

%The system begins to evolve from a regular grid lattice with the size $L=50$, 
Let us consider a regular grid lattice of size $L$
(the total number of players is $N=L\times L$), 
which satisfies the periodic boundary conditions, and each node of the lattice will be occupied by a player who has $8$ nearest neighbors (\textit{i.e.}, we consider the Moore neighborhood). 
Initially, each player will be randomly assigned 
equiprobably to one of three possible strategies: Cooperator (ALLC), Defector (ALLD) or Discriminator (DISC), which can be described in detail as follows:
\begin{itemize}
  \item \textbf{ALLC}: the donor always cooperates, that is, ALLC strategists are unconditional cooperators.
  \item \textbf{ALLD}: the donor defects under any scenario. ALLD strategists are unconditional defectors.
  \item \textbf{DISC}: the decision of whether to cooperate or not depends on the estimated reputation score of the recipient, which in turn is based i) on the last recipient's actions and ii) on the donor's assessment rule. We term these strategists as discriminators; the assessment rules have been described in previous Section \ref{Subsection:Second-orderAssesmentRules}.
\end{itemize}

Players, as donors, are characterized by four possible actions: CC, CD, DC, DD. Two of these actions, CC and CD, correspond to cooperative actions, namely, CC when cooperating against a cooperator (\textit{i.e.}, the recipient cooperated in his last action) and CD when cooperating against a defector (recipient's last action was to defect). The other two actions, DC and DD, correspond to non-cooperative actions: DC if a player defects against a cooperator and DD if he defects against a defector. In order to characterize the memory effect of individuals in the current model, we will record the action lists for each individual in the most recent $M$ steps \cite{cuest15_sp}.

%, and $M$ is usually set to be $5$ without lacking the generality since Cuesta \textit{et al.}
% found that the longer memory does not favor the evolution of cooperation. 

Regarding the initial conditions, first $M$ actions of ALLC (\textit{resp.}, ALLD) strategists are randomly chosen from CC or CD (\textit{resp.}, DC or DD), while first $M$ actions of DISC strategists 
are randomly taken from the set \{CC, CD, DC, DD\} and determined by the specific assessment rule of the discriminator (%will be displayed in
  Section \ref{Subsection:Second-orderAssesmentRules}).

%Regarding the initial conditions, first $M$ actions in the memory history for any strategist are initialized according to the following cases:
%\begin{itemize}
%  \item \textbf{ALLC strategists}: their initial $M$ actions are randomly chosen from CC or CD.%, which means that the cooperative donor contributes to a cooperator or to a defector, respectively.%  recipient with a good or bad image;
%  \item \textbf{ALLD  strategists}: initial actions are stochastically selected from DC or DD.%, which implies the donor rejects the donation to his cooperator or defector partner, respectively.
%  \item \textbf{DISC strategists}: initial actions randomly taken from a set \{CC, CD, DC, DD\} and determined by the specific assessment rule of the discriminator (%will be displayed in
%Section \ref{Subsection:Second-orderAssesmentRules}).
%\end{itemize}

\subsection{Iteration Procedure}

%After that, the system iterates forwards and one independent simulation is made up of $g=2000$ generations which ensure that the evolution can arrive at the steady state. 

The evolution of the game will be hinged in the following way: at each elementary time step, a random player (the focal player or donor) chooses a random neighbor (the recipient) and decides if he cooperates or not. At each period, any player will have, on average, one chance to act as a donor, that is, a period consists of $N$ elementary time steps (Donation Game decisions) that will take part in random order. 

%In turn, each generation includes $h$ periods within the payoff is accumulated for the strategy update at the end of the generation. Players' accumulated payoffs counters will be reset at the beginning of each generation. 

%$h=50$
%Regarding the memory of individual actions, we will also synchronously copy the action lists from the imitated neighbors to the focal ones. On the contrary, we can also only update the strategies, and keep their original action lists for all individuals as a reference.

%At the end of a generation, all individuals in the population will synchronously update their strategies. 
%A period is composed of two elementary phases as follows,
%\begin{description}
%  \item[1)]\textbf{Player-selection phase}. As mentioned above, one player (e.g., Player-$i$) can be randomly chosen as a donor and will have an opportunity to perform the donation game with a recipient (say, Player-$j$), who is also stochastically selected from $8$ nearest neighbors of Player-$i$.
%  \item[2)]\textbf{Donation-game phase}. One-shot donation game may be conducted, and whether this game will actually happen is dependent upon the strategy of Player-$i$, that is,  the focal Player-$i$ will decide to make the donation according to the specific cases (whose details will be provided later).
%\end{description}

%For the sake of clarity, let us explain in detail the dynamical procedure:
Let us explain in detail the dynamical procedure:

%Thus, the evolution of game will be hinged upon the strategy of Player-$i$ in the following way,
\begin{itemize}
  \item \textbf{ALLC}: if the focal Player-$i$ is an unconditional cooperator, he pays the cost $c$ to Player-$j$ who obtains the benefit $b$ ($i$ and $j$ payoffs are $-c$ and $b$, respectively). We record $i$'s last action as CC if Player-$j$'s last action was CC or CD; otherwise, the last action of Player-$i$ is recorded as CD.
  \item \textbf{ALLD}: if the focal Player-$i$ is an unconditional defector, he rejects the donation to his partner $j$, and both payoffs are zero. We record $i$'s last action as DC if Player-$j$'s last action was CC or CD; otherwise, the last action of Player-$i$ is recorded as DD.
  \item \textbf{DISC}: if the focal Player-$i$ is a discriminator, he will calculate the weighted image score of Player-$j$ in the light of four different assessment rules as shown in Tab. \ref{tab1}. If Player-$j$'s score is higher than the required minimum reputation $H_0$, Player-$i$ will donate to $j$; otherwise, Player-$i$ will reject the donation.
  $H_0$ represents a minimum threshold so that the recipient can be considered good enough to be a beneficiary of the donation. It is, therefore, a measure of the intolerance. Finally, Player-$i$'s last action will be accordingly updated. The detailed decision procedure for DISC players can be further described as follows:

\begin{description}
  \item[1)]\textbf{Assessment}.%\textbf{Judgement of the goodness}.
  Player-$i$ will evaluate Player-$j$'s actions to be good (G) or bad (B) according to Tab. \ref{tab1}. % on the basis of the action list of Player-$j$.
  As an example, we assume that Player-$i$ is a discriminator adopting the Stern Judging rule, and the 
  %action list of Player-$j$ (i.e., Player-$j$'s memory) 
  %has been recorded as {CC, DC, CD, DD, CC}. 
  last $M=5$ actions of Player-$j$ are {CC, DC, CD, DD, CC}.
  Then, based on Tab. \ref{tab1}, Player-$i$ judges Player-$j$'s goodness of action list to be {G, B, B, G, G}.
  \item[2)]\textbf{Calculation of the weighted score}. If the action is judged as good (G) or bad (B), the corresponding score will be $1$ or $0$, respectively. The final reputation score of Player-$j$ through the eyes of Player-$i$ will be defined as:
\begin{equation}
r_{j|i}=w*S+(1-w)*\bar{S}\;,
\label{reputationScore}
\end{equation}
where $S$ denotes the score of $j$'s last action and $\bar{S}$ represents the average score of $j$'s 
%whole action list 
$M$ last actions
%including the last action
\footnote{Cuesta \textit{et al.} \cite{cuest15_sp} showed through lab-based human experiments that people measures reputation based on the weighted average of the fraction of cooperative actions ($\bar{C}$) and the last action performed ($C_{last}$), in which the coupling relationship can be linearly characterized as $w*C_{last}+(1-w)*\bar{C}$ and $w$ is often fitted to be $0.165$ from the experimental data.}. In the above-mentioned example, Player-$j$'s score will be: $r_{j|i}=w*1+(1-w)*\frac{1+0+0+1+1}{5}=w+(1-w)(3/5)$.%=(2W+3)/5$
  \item[3)]\textbf{Decision}. %After computing the reputation score of Player-$j$, Player-$i$ compares $r_{j|i}$ with $H_0$. 
  If $r_{j|i}>H_0$, Player-$i$ will pay the cost $c$ to cooperate with Player-$j$, who will obtain the benefit $b$. Otherwise,
  Player-$i$ will defect, and both payoffs will be zero. We will record
  Player-$i$ last action accordingly.
  
  %After computing the reputation score of Player-$j$, Player-$i$ compares $r_{j|i}$ with $H_0$. If $r_{j|i}>H_0$, Player-$i$ will decide to pay the cost $c$ to cooperate with $j$ (who will obtain the benefit $b$) and update his last action as CC if Player-$j$'s last action was CC or CD, or CD if $j$'s last action was DC or DD. On the contrary, if $r_{j|i}\le H_0$, Player-$i$ will reject the donation and pay nothing, meanwhile Player-$j$ will also get nothing. Similarly, Player-$i$ will record his last action as DC if Player-$j$'s last action was CC or CD, or DD if $j$'s last action was DC or DD.
\end{description}
\end{itemize}

%For a discriminator (e.g., Player-$i$), whether he will donate is determined by the image score of his opponent (e.g., Player-$j$) and 

A generation includes $h$ of the above described periods. 
At the end of a generation, all the players synchronously update their current strategies following a Fermi-like updating rule \cite{fermi98,diversity2,diversity3}.
Let $P_i$ and $P_j$ be the payoffs of player $i$ and a random neighbor $j$,
accumulated throughout the last generation. Then, Player-$i$ will imitate Player-$j$'s strategy with a probability $Prob(i\leftarrow j)$ given by:
\begin{equation}
%Prob(i\leftarrow j)=\frac{1}{1+e^{\frac{-(P_j-P_i)}{K}}}\;,
Prob(i\leftarrow j)=\frac{1}{1+e^{(P_i-P_j)/K}}\;,
\end{equation}
where $K$ denotes the irrationality of individual choice or the noise of strategy adoption. %In the case of imitation, both strategy and action's list are copied from the imitated player to the focal one.

Note that the model includes two different time scales: a time scale involving payoff-independent decision-making strategies \cite{yamir2012_pnas,gracia12_SciRep}, and another longer scale, of evolutionary character, involving strategies imitation \cite{fermi98}.

\section{Mean-field approximation.}
\label{section:MeanField}

In this Section, we discuss various approaches to obtain a mean-field solution to the model here presented. These approaches preserve the assessment rules based on the second-order reputation while neglecting some aspects related to the spacial distribution and formation of clusters, the length of the memory and the weight of the last action. The goal of this mean-field approximation is to capture  the qualitative behavior of the system and detect which specific aspects are not reproduced due to the ingredients not considered here.
Throughout this section, we will refer to figures \ref{fig:1} and \ref{fig:2} (which contain the numerical results that will be developed in Section \ref{Section:Simulations}) to have a visual reference of the parameter space and, also, to compare the predictions with the agent-based numerical results.
 
Consider a well-mixed population and the low noise case ($K\ll hb$), which allows us to assume a deterministic imitation rule. Let $\rho_{c}$, $\rho_{d}$, and $\rho_{i}$ be the fraction of ALLC, ALLD, and DISC strategists, respectively.
%In order to be consistent with the numerical simulations, we 
For simplicity, let us  
consider an initial population defined by $\rho_{c}=\rho_{d}=\rho_{i}=1/3$.
%we consider an initial population composed of 1/3 of unconditional ALLC, 1/3 of unconditional ALLD, and 1/3 of DISC strategists.

\subsection{Image Scoring} 
Here, we study the case when DISCs are Image Scoring strategists. Discriminators always donate to ALLC, but never to ALLD players. Let $\langle r_{i|i} \rangle$ be the mean value of the reputation score of a DISC as seen by another DISC. From Eq. (\ref{reputationScore}) and Table \ref{tab1}, it follows:

\begin{equation}
\langle r_{i|i} \rangle = w \langle C \rangle + (1-w) \langle C \rangle = \langle C \rangle,
\end{equation}
where  $\langle C \rangle$ is the fraction of cooperative actions in the system. Note that $w$ influences the variance of $r_{i|i}$ but not its mean value. On average, a DISC will give to another DISC if $\langle r_{i|i} \rangle>H_0$.\\

The average payoffs for ALLC, ALLD, and DISC strategists are, respectively:

\begin{eqnarray}
\Pi_c&=&(\rho_c+\rho_i)b-c\nonumber\\
\Pi_d&=&\rho_cb\nonumber\\
\Pi_i&=&(\rho_c+\rho_i\Theta(\langle r_{i|i} \rangle-H_0))(b-c)\nonumber\\
&=&(\rho_c+\rho_i\Theta(\langle C \rangle-H_0))(b-c), \nonumber\\ \label{PiIS}
\end{eqnarray}
where $\Theta$ stands for the Heaviside function which is zero for negative arguments and one for positive ones.

\subsubsection*{Low $H_0$} For low enough values of $H_0$ (\textit{i.e.}, $H_0 < \langle C \rangle$), DISC players, on average, will cooperate when facing another DISC. The average cooperation level within a generation (constant $\rho_c,\rho_d,\rho_i$) evolves according to:

\begin{equation}
\langle C \rangle \gtrsim \rho_{c}+\rho_{i}\langle C \rangle \;\; \rightarrow \;\; \langle C \rangle \gtrsim \frac{\rho_c}{1-\rho_i}. \\
\label{c_ISlowH}
\end{equation}

Within the first generation, given $\rho_c=\rho_d=\rho_i=1/3$, $\langle C \rangle$ evolves to a value greater than $1/2$. When $H_0 < \langle C \rangle$, what is true %in the early stages 
for $H_0 \lesssim 0.5$, the average payoff of a DISC is given by:

\begin{equation}
\Pi_i=(\rho_c+\rho_i)(b-c), \label{Pi_iISlowH}
\end{equation}
and the average payoff difference between DISC and ALLD is:
\begin{equation}
\Pi_i-\Pi_d=(\rho_c+\rho_i)(b-c)-\rho_cb,
\end{equation}
what implies that $\Pi_i>\Pi_d$ if $\rho_ib>(\rho_c+\rho_i)c$. At the end of the first generation, this condition will be satisfied for $b>2c$. For $b>2c$, DISC will overcome both ALLD and ALLC, while ALLD will overcome ALLC. If $\langle C \rangle$ increases over time, condition $H_0 < \langle C \rangle$ is preserved, and therefore Eq. (\ref{Pi_iISlowH}). Furthermore, as $\rho_i$ increases over time, $\Pi_i-\Pi_d$ increases. So, we can conclude that,  for $H_0 \lesssim 0.5$, DISC will invade ALLD if $b>2c$.

Regarding the resilience of ALLC strategists, on the one hand the average payoff difference between ALLC and ALLD is given by:
\begin{equation}
\Pi_c-\Pi_d=\rho_ib-c,
\end{equation}\label{c-d_IS}
what implies that $\Pi_c>\Pi_d$ if $\rho_ib>c$. For the initial strategists distribution ($\rho_i=1/3$), ALLC defeats ALLD for $b>3c$. On the other hand, the payoff difference between DISC and ALLC is:\begin{equation}
\Pi_i-\Pi_c=c\rho_d,
\end{equation}
that is, $\Pi_i>\Pi_c$ if $\rho_d>0$, $\Pi_i=\Pi_c$ otherwise. Actually, in absence of ALLD players, ALLC and DISC are indistinguishable strategists.

Summarizing, regarding panels (III) in Fig. \ref{fig:2}, where $c=1$:

\begin{itemize}

\item Upper left area. Provided $b>3$, the higher the value of $b$, the higher the fraction of ALLC that will survive the first stages and will coexist with DISC players at the steady state. 

\item For $3>b>2$, DISC will invade ALLC and ALLD.

\item Bottom left corner. For $b<2$, we have $\Pi_d-\Pi_i=\rho_c$, and
ALLD will invade ALLC and DISC.

\end{itemize}

\subsubsection*{High $H_0$} For high values of $H_0$ (\textit{i.e.}, $H_0 < \langle C \rangle$), the average cooperation level within a generation (constant $\rho_c,\rho_d,\rho_i$) evolves according to:

\begin{equation}
\langle C \rangle \lesssim \rho_{c}+\rho_{i}\langle C \rangle \;\; \rightarrow \;\; \langle C \rangle \lesssim \frac{\rho_c}{1-\rho_i}. \\
\label{c_IS_HighH}
\end{equation}

Within the first generation ($\rho_c=\rho_d=\rho_i=1/3$), $\langle C \rangle$ evolves to a value lower than $1/2$. When $H_0>0.5$, the average payoff of a DISC is given by:
\begin{equation}
\Pi_i=\rho_c(b-c),
\end{equation}
and, therefore:
\begin{eqnarray}
\Pi_c-\Pi_i&=&\rho_dc+\rho_i(b-c),\nonumber\\
\Pi_d-\Pi_i&=&\rho_cc.\nonumber\\
\end{eqnarray}

For high enough values of $b$, both ALLC and ALLD will defeat DISC. Taking into account:
\begin{equation}
\Pi_d-\Pi_c=c-\rho_ib
\end{equation}
the advantage of the ALLD over the ALLC increases as $\rho_i$ decreases, which in turn leads to a reduction in $\rho_c$ and to an absorving mono-strategic state of ALLD (Upper-right and central-right area of panels (III) in Fig. \ref{fig:2}).

As $b$ decreases, the average payoff difference between ALLC and DISC %($\rho_dc+\rho_i(b-c)=\rho_ib-(\rho_i-\rho_d)c$)
(i.e., $\rho_ib-(\rho_i-\rho_d)c$) 
decreases. Note that, in absence of ALLC, DISC strategists never donate, therefore ALLD and DISC are indistinguishable strategists: if some DISC strategists survive the first stages sorrounded by DISC and ALLD, they will coexist (as defectors) with ALLD, allowing a mixed equilibrium of ALLD and DISC (Bottom-right area of panels (III) in Fig. \ref{fig:2}). Note that, in any case, the only action will be to defect (right area of panel (c) in Fig. \ref{fig:1}).

\subsection{\textbf{Shunning}} 

In this subsection, we analize the case when DISCs are Shunning strategists. In this case, discriminators never donate to ALLD. On the other hand, ALLC always cooperate, but only when cooperating with a cooperator will be positively evaluated by DISC. %According to that% and Eq. (\ref{reputationScore})
%, the mean value of the reputation score of an ALLC as seen by a DISC is given by:

The mean reputation scores of ALLC, ALLD, and DISC players through the eyes of a DISC are, respectively:

%\langle r_{c|i} \rangle = w \langle C \rangle + (1-w) \langle C \rangle = \langle C \rangle.
\begin{eqnarray}
\langle r_{c|i} \rangle &=& \langle C \rangle,\nonumber\\
\langle r_{d|i} \rangle &=& 0,\nonumber\\
%\langle r_{i|i} \rangle &=& \rho_c \langle C \rangle + \rho_i \langle C \rangle^2,\nonumber\\
\langle r_{i|i} \rangle &=&  \langle C \rangle^2.\nonumber\\
\label{r_Sh}
\end{eqnarray}
%where  $\langle C \rangle$ is the fraction of cooperative actions in the system.% On average, a DISC will give to an ALLC if $\langle r_{c|i} \rangle>H_0$, and to another DISC if $\langle r_{i|i} \rangle>H_0$.\\

\subsubsection*{Low $H_0$}

For low relative values of $H_0$ (\textit{i.e.}, $H_0 < \langle C \rangle$), DISC players, on average, will pay to ALLC. %Now we have:
%\begin{equation}
%\langle r_{i|i} \rangle = (\rho_c+ \rho_i) \langle C \rangle.
%\end{equation}
%and, therefore, for $H_0<(\rho_c+ \rho_i) \langle C \rangle$, DISC players will pay to ALLC and DISC. 
The average cooperation level within a generation (constant $\rho_c,\rho_d,\rho_i$) evolves according to:

\begin{equation}
\langle C \rangle \gtrsim \rho_{c}+\rho_{i}\langle C \rangle^2  \;\; \rightarrow \;\; \langle C \rangle \gtrsim \frac{1-\sqrt{1-4\rho_c\rho_i}}{2\rho_i}. \\
\label{c_SH}
\end{equation}
In the first generation ($\rho_c=\rho_d=\rho_i=1/3$), $\langle C \rangle$ evolves to %$(3 - \sqrt(5))/2\simeq 0.38$.
a value greater than $\sim 0.38$.
When $H_0<\langle C \rangle^2$, what is true in the early stages for $H_0 \lesssim 0.15$, the average payoffs are given by:
\begin{eqnarray}
\Pi_c&=&(\rho_c+\rho_i)b-c,\nonumber\\
\Pi_d&=&\rho_cb,\nonumber\\
\Pi_i&=&(\rho_c+\rho_i)(b-c)\nonumber\\ \label{payoffsShunningLowH}
\end{eqnarray}
which are the same payoffs that those of Eq. (\ref{PiIS},\ref{Pi_iISlowH}) corresponding to the previous Image Scoring - Low $H_0$ case, and therefore, the same analysis applies here. Note that, although payoffs in Eq. (\ref{payoffsShunningLowH}) were calculated for the first generation, $\langle C \rangle$ increases over time, and therefore also the payoffs differences. A consequence is the similarity between the left part of the respective panels (I) and (III) in Fig. \ref{fig:2}, and between panels (a) and (c) in Fig. \ref{fig:1}. Nevertheless, given the fact that the condition of low $H_0$ is more restrictive in the current case, the cooperative green area on the right side of panel (a) in Fig. \ref{fig:1} is smaller than that in panel (c).

\subsubsection*{High $H_0$}

For high values of $H_0$ (\textit{i.e.}, $H_0 > \langle C \rangle$), DISC players, on average, will not donate to ALLC. The cooperation level within a generation %(constant $\rho_c,\rho_d,\rho_i$) 
evolves according to:

\begin{equation}
\langle C \rangle \lesssim \rho_{c}+\rho_{i}\langle C \rangle^2  \;\; \rightarrow \;\; \langle C \rangle \lesssim \frac{1-\sqrt{1-4\rho_c\rho_i}}{2\rho_i}. \\
\label{c_SH}
\end{equation}

Therefore, the condition $H_0>\langle r_{i|i} \rangle = \langle C \rangle^2$ is
satisfied at the end of the first generation ($\rho_c=\rho_i=1/3$, $\langle C \rangle \lesssim 0.38$). %for  $\langle C \rangle$. As $\rho_i \leq 1$, condition (\ref{R_i_SH_HighH}) is always satisfied in this subcase $H_0 > \langle C \rangle$, what at the end of the first generation it is true for $\langle C \rangle > \sim 0.38$).
%Therefore, we can distinguish two cases:
%\begin{eqnarray}
%i) H_0>\rho_c \langle C \rangle\nonumber \\
%ii) H_0>\rho_c \langle C \rangle\nonumber \\
%\end{eqnarray}\nonumber
It follows that DISC players, on average, will not donate to anybody. The payoffs are given by:
\begin{eqnarray}
\Pi_c&=&\rho_cb-c,\nonumber\\
\Pi_d\;=\;\Pi_i&=&\rho_cb.
\label{payoffShunningHighC}
\end{eqnarray}
%Consequently
For high values of $H_0$, DISC will play as ALLD, both having a higher payoff than ALLC.

Payoffs in (\ref{payoffShunningHighC}) were calculated for the first generation. Nevertheless, $\langle C \rangle$ decreases over time as DISC and ALLD overcome ALLC, and therefore $\langle r_{i|i}\rangle$ decreases according to (\ref{r_Sh}). This means that the order of the payoffs is preserved over time: ALLD and DISC will invade ALLC. Although there is a mixed equilibrium composed of DISC and ALLD strategists (right side of panels (I) in Fig. \ref{fig:2}), the former will act as defectors, and therefore the cooperative level will tend to zero (right side of panel (a) in Fig. \ref{fig:1}).

Since in this case $\Pi_d=\Pi_i$, higher-order effects 
beyond the mean-field approach should play a key role. Although for $H_0 > \langle C \rangle$, DISC players, on average, will not donate to any strategist, %they have an $\epsilon>0$ probablity of paying to an ALLC 
there is an $\epsilon>0$ probablity for a DISC to pay to an ALLC 
(and an even smaller probability to pay to another DISC). By adding this corrective term, the average DISC payoff becomes:

\begin{equation}
\Pi_i=\rho_cb-\epsilon\rho_cc=\rho_c(b-\epsilon c).
\end{equation}
The lower the value of $H_0$ (also the shorter the memory length $M$), the higher the value of $\epsilon$. Furthermore, the relative payoff difference between 
ALLD and DISC will decrease as $b$ increases. To summarize, although according to the mean-field approximation ALLD and DISC payoffs are equal, higher-order effects imply a dependence of the final mixed equilibrium on
$b$ and $H_0$.

\subsection{\textbf{Stern Judging}} 

Here, we investigate the case when DISCs are Stern Judging strategists. Depending on the values of the parameters, DISC players may donate or not both to ALLC and ALLD players. The mean reputation scores of the different strategist, as seen by a DISC, are given by:

%\langle r_{c|i} \rangle &=& w \langle C \rangle + (1-w) \langle C \rangle = \langle C \rangle,\nonumber\\
%\langle r_{d|i} \rangle &=& w (1- \langle C \rangle )+ (1-w) (1- \langle C \rangle ) = (1- \langle C \rangle )\nonumber\\
\begin{eqnarray}
\langle r_{c|i} \rangle &=& \langle C \rangle,\nonumber\\
\langle r_{d|i} \rangle &=& 1- \langle C \rangle,\nonumber\\
\langle r_{i|i} \rangle &=& \langle C \rangle^2 + (1- \langle C \rangle)^2.\nonumber\\
%\langle r_{i|i} \rangle &=& \rho_c \langle C \rangle + \rho_d (1- \langle C \rangle) + \rho_i (\langle C \rangle^2 + (1- \langle C \rangle)^2).\nonumber\\
\label{rSJ}
\end{eqnarray}
%where  $\langle C \rangle$ is the fraction of cooperative actions in the system.

\subsubsection*{Low $H_0$}

%Note that we are not considering the variance of reputations ($r_{c|i},r_{d|i},r_{i|i}$). Actually, although for $H_0 < 1/2$, DISC players, on average, will evaluate positively any strategist, the shorter the memory length $M$, the higher the probability $\epsilon$ that a DISC will defect. The same applies to $H_O$: the higher the value of $H_O$, the higher the fraction of defection among DISC strategists. 

From (\ref{rSJ}) it follows that, provided $H_0 < \langle C \rangle$ and $H_0 < (1-\langle C \rangle$), DISC players, on average, will pay to ALLC and ALLD. At the first generation ($\rho_c=\rho_d=\rho_i=1/3$), on average, half of the actions of a DISC will be considered as good actions by another DISC. The average cooperation level within a generation (constant $\rho_c, \rho_d , \rho_i$) evolves according to:
 
\begin{eqnarray}
\langle C \rangle \gtrsim \rho_c +\rho_i \langle C \rangle^2+\rho_i(1-\langle C \rangle)^2 \rightarrow \nonumber\\
\nonumber\\
\langle C \rangle \gtrsim \frac{2\rho_i+1-\sqrt{1-4\rho_i^2-8\rho_c\rho_i+4\rho_i}}{4\rho_i}
\label{cSJ}
\end{eqnarray}

Therefore, for %$H_0 < \langle 1/3 \rangle$ and 
($\rho_c=\rho_d=\rho_i=1/3$), it follows $\langle r_{i|i} \rangle \gtrsim  1/2$.
The corresponding average payoffs for $H_0 < 1/3$ will be:

\begin{eqnarray}
\Pi_c \; = \; \Pi_i&=&(\rho_c+\rho_i)b-c,\nonumber\\
\Pi_d&=&(\rho_c+\rho_i)b.\nonumber\\
%\Pi_i&=&(\rho_c+\rho_i)b-c,\nonumber\\
\end{eqnarray}
%As ALLD players have the highest average payoff, ALLD strategy will invade ALLC and DISC (left area of panels (II) in Fig. \ref{fig:2}). 

At the end of the first generation, ALLD players overcome ALLC and DISC and $\langle C \rangle$ decreases. Therefore, according to (\ref{rSJ}), $\langle r_{c|i}\rangle$ decreases and $\langle r_{d|i} \rangle$ increases over time. In the same way, $\langle r_{i|i}\rangle$ tends to $\rho_d + \rho_i$ as $\langle C \rangle$ decreases, and therefore to 1 as $\rho_c$ decreases. Consequently, payoffs evolve over time towards:

\begin{eqnarray}
\Pi_c &=& \rho_cb-c,\nonumber\\
\Pi_d &=& (\rho_c+\rho_i)b,\nonumber\\
\Pi_i &=& (\rho_c+\rho_i)b-c,\nonumber\\
\label{PiSJLowH}
\end{eqnarray}
and  ALLD strategy will invade ALLC and DISC (left area of panels (II) in Fig. \ref{fig:2}). 

% &=& 1- \langle C \rangle 
%\nonumber\\
%\langle r_{i|i} \rangle &=& \rho_c \langle C \rangle + \rho_d (1- \langle C \rangle) + \rho_i (\langle C %\rangle^2 + (1- \langle C \rangle)^2)\nonumber\\

%have the highest average payoff,

Mean-field approximation cannot reproduce the cooperative behavior observed in the numerical simulations for high values of $b$, when payoff differences are small and other high-order %factors such as the topology of the network 
effects become key. As in the previous case, there are two equal payoffs in (\ref{PiSJLowH}), here $\Pi_c=\Pi_i$. By adding a higher-order corrective term, the average DISC payoff for the first stages becomes:
\begin{equation}
\Pi_i=(\rho_c+\rho_i)b-(1-\epsilon)c.\nonumber\\
\end{equation}
Note that, unlike the previous case (Shunning, high $H_0$), the corrective term now applies to the probability of a DISC to defect against any strategist (\textit{i.e.}, it is not multiplied by a density $\rho$), becoming higher than that for Shunning discriminators. The relative payoff difference between 
ALLD and the rest of the players will decrease as $b$ increases. For high values of $b$, the differences between the payoffs cannot prevent the formation of cooperative clusters. Note that this cooperative behavior (upper left corner of panel (b) in Fig. \ref{fig:1}) corresponds to DISC strategist that act as cooperators.

\subsubsection*{High $H_0$}

For high relative values of $H_0$ (\textit{i.e.}, $H_0 > \langle C \rangle$ and $H_0 > (1-\langle C \rangle$), DISC players, on average, will pay neither ALLC nor ALLD. As in the previous case (low $H_0$), at the first generation ($\rho_c=\rho_d=\rho_i=1/3$), on average, half of the actions of a DISC will be considered as good actions by another DISC. The average cooperation level within a generation
evolves according to:
 
\begin{eqnarray}
\langle C \rangle \lesssim \rho_c +\rho_i \langle C \rangle^2+\rho_i(1-\langle C \rangle)^2 \rightarrow \nonumber\\
\nonumber\\
\langle C \rangle \lesssim \frac{2\rho_i+1-\sqrt{1-4\rho_i^2-8\rho_c\rho_i+4\rho_i}}{4\rho_i}
\label{cSJHighH}
\end{eqnarray}
Solving it for $\rho_c=\rho_i=1/3$, it is found that for $H_0>1/2$, a DISC will probably defect when facing any strategist.
Regarding higher order effects, the shorter the memory length $M$, the higher the probability $\epsilon$ for a DISC to cooperate.

At the end of the first generation, the corresponding average payoffs will be:

\begin{eqnarray}
\Pi_c&=&\rho_cb-c,\nonumber\\
\Pi_d&=&\rho_cb,\nonumber\\
\Pi_i&=&\rho_cb-\epsilon c.\nonumber\\
\end{eqnarray}
where, the higher-order corrective term $\epsilon$ has been added ($\Pi_i=\Pi_d$ in the mean-field). Given $\Pi_d>\Pi_i>\Pi_c$, ALLD players will beat ALLC and DISC. 
%Therefore, according to (\ref{rSJ}), as $\langle C \rangle$ decreases  over time, $\langle r_{c|i}\rangle$ decreases, while $\langle r_{d|i} \rangle$ and $\langle r_{i|i}\rangle$ increase. 
%According to (\ref{rSJ}), as $\langle C \rangle$ decreases over time, $\langle r_{c|i}\rangle$ also decreases, while $\langle r_{d|i} \rangle$ and $\langle r_{i|i}\rangle$ increase. 
According to (\ref{rSJ}), the consequent decrease of $\langle C \rangle$ leads to an increase in $\langle r_{d|i} \rangle$ and $\langle r_{i|i}\rangle$, and to a decrease in $\langle r_{c|i}\rangle$.

Consequently, payoffs evolve over time towards:

\begin{eqnarray}
\Pi_c &=& \rho_cb-c,\nonumber\\
\Pi_d &=& (\rho_c+\rho_i)b,\nonumber\\
\Pi_i &=& (\rho_c+\rho_i)b-c,\nonumber\\
\end{eqnarray}
and  ALLD strategy will invade ALLC and DISC, bringing the system to a mono-strategic ALLD state 
%the system will reach an absorbing mono-strategic ALLD state 
(right area of panels (II) in Fig. \ref{fig:2}).

\subsection{\textbf{Simple Standing}} 

In this subsection we discuss the case when DISC are Simple Standing strategists. In this case, discriminators always cooperate when facing an ALLC. Regarding ALLD strategists, an ALLD defecting against a defector will be positively evaluated by a DISC. %According to that, the mean value of the reputation score of an ALLD as seen by a DISC is given by:
The mean value of the reputation scores through the eyes of a DISC are:\\

\begin{eqnarray}
\langle r_{c|i} \rangle &=& 1,\nonumber\\
\langle r_{d|i} \rangle &=& 1-\langle C \rangle,\nonumber\\
\langle r_{i|i} \rangle &=& \langle C \rangle+(1-\langle C \rangle)^2.\nonumber\\
%\langle r_{i|i} \rangle &=& \rho_c + \rho_d (1-\langle C \rangle) +  \rho_i  (\langle C \rangle+(1-\langle C \rangle)^2).\nonumber\\
\label{rSS}
\end{eqnarray}
%where  $\langle C \rangle$ is the fraction of cooperative actions in the system.

On average, a DISC will give to an ALLD if ${\langle r_{d|i} \rangle =  1 - \langle C \rangle > H_0}$. Regarding how a DISC evaluates another DISC, note that the function $\langle r_{i|i}\rangle (\langle C \rangle)$ is not monotonous, reaching a minimun value for $\langle C \rangle=1/2$. 

%\subsubsection*{Low $H_0$}

For low relative values of $H_0$ (\textit{i.e.}, $H_0 < 1- \langle C \rangle$), DISC players, on average, will pay to ALLD. Within a generation, the average cooperation evolves according to:

\begin{equation}
\langle C \rangle \gtrsim \rho_{c}+\rho_{i}(\langle C \rangle+(1-\langle C \rangle)^2). \\
\label{c_SS}
\end{equation}

Solving (\ref{c_SS}) for $\rho_c=\rho_i=1/3$, it is found that $\langle C \rangle$ %evolves to a value higher or equal to $2-\sqrt{2}\sim0.59$. 
evolves towards $\langle C \rangle \gtrsim 0.59$ within the first generation.
It follows that, at the end of the first generation, DISC will pay to all the strategists for $H\lesssim 0.41$. The corresponding average payoffs are given by:

\begin{eqnarray}
\Pi_c\;=\;\Pi_i&=&(\rho_c+\rho_i)b-c,\nonumber\\
\Pi_d&=&(\rho_c+\rho_i)b.\nonumber\\
\label{payoffsSSLowH}
\end{eqnarray}
Therefore, ALLD players will overcome ALLC and DISC. As $\rho_d$ and ($1-\langle C \rangle$) increase over time, $r_{i|i}$ increases, and the system will evolve in time towards $\langle r_{c|i}\rangle = \langle r_{d|i}\rangle = \langle r_{i|i} \rangle = 1$, with DISC playing as ALLC. The system is characterized by a fraction $\rho_c+\rho_d$ of cooperators and a fraction $\rho_d$ of defectors. This is the classical scenario where
mean-field approach involves full defection (ALLD) and cannot explain cooperation for high enough values of $b$ in structured populations (and also with memory in this model).

%\subsubsection*{High $H_0$}

As $H_0$ increases, the probablity for a DISC to pay to an ALLD decreases. For $H_0 > 1- \langle C \rangle$, DISC players, on average, will not donate to ALLD. Approximation (\ref{c_SS}) becomes:
\begin{equation}
\langle C \rangle \lesssim \rho_{c}+\rho_{i}(\langle C \rangle+(1-\langle C \rangle)^2). \\
\label{c_SS-HighH}
\end{equation}
Solving (\ref{c_SS-HighH}) for $\rho_{c}=\rho_{i}=1/3$, it is found that within the first generation the cooperation will tend to $\langle C \rangle\rightarrow\sim 0.59$. At the end of the first generation, for $H_0 \gtrsim 0.41$, and the average payoffs can be approximated by:
\begin{eqnarray}
\Pi_c&=&(\rho_c+\rho_i)b-c,\nonumber\\
\Pi_d&=&\rho_cb.\nonumber\\
\Pi_i&=&(\rho_c+\rho_i)(b-c),\nonumber\\
\label{payoffsSSHighH}
\end{eqnarray}
and the average payoff difference between DISC and ALLD will be:
\begin{equation}
\Pi_i-\Pi_d=(\rho_c+\rho_i)(b-c)-\rho_cb=\rho_ib-(\rho_c+\rho_i)c.
\end{equation}
%what implies that $\Pi_i>\Pi_d$ if $\rho_i(b-c)>\rho_cb$.
%Therefore, $\rho_ib>(\rho_c+\rho_i)c$ implies $\Pi_i>\Pi_d$. 
In the first stages ($\rho_c=\rho_i$), DISC players will overcome ALLC and ALLD for $b>2c$.
%Nevertheless, as $\rho_i$ and $\langle C \rangle$ increases over time, $\langle r_{d|i}\rangle$ increases, and $\langle r_{i|i}\rangle$ decreases. 
Nevertheless, the consequent increase of $\rho_i$ and $\langle C \rangle$ over time leads to a decrease in
$\langle r_{i|i}\rangle$ and to an increase in $\langle r_{d|i}\rangle$. This fact involves a trade-off between $b$ and $H_0$: %provided $b>2c$, 
a lower value of $H_0$ involves a higher $b$ to allow DISC invading ALLD (column (IV) in Fig. \ref{fig:2}).

Regarding ALLC strategists, the average payoffs differences are given by:
\begin{eqnarray}
\Pi_c-\Pi_d=\rho_ib-c,\nonumber\\
\Pi_i-\Pi_c=c\rho_d,\nonumber\\
\end{eqnarray}
what implies that $\Pi_c>\Pi_d$ for $\rho_ib>c$. At the first stages ($\rho_i=1/3$), ALLC defeats ALLD for $b>3c$. Furthermore, $\Pi_i>\Pi_c$ if $\rho_d>0$, otherwise $\Pi_i=\Pi_c$. Actually, for $\rho_d=0$, ALLC and DISC are indistinguishable strategists. Provided $b \geq 3$, the higher the value of $b$, the higher the fraction of ALLC players that will survive the first stages and will coexist with DISC ones at the steady state (panels (d1,d3) in Fig. \ref{fig:2}).

\section{Numerical simulations}
\label{Section:Simulations}

\begin{figure}
\begin{center}
\includegraphics[width=1.0\columnwidth]{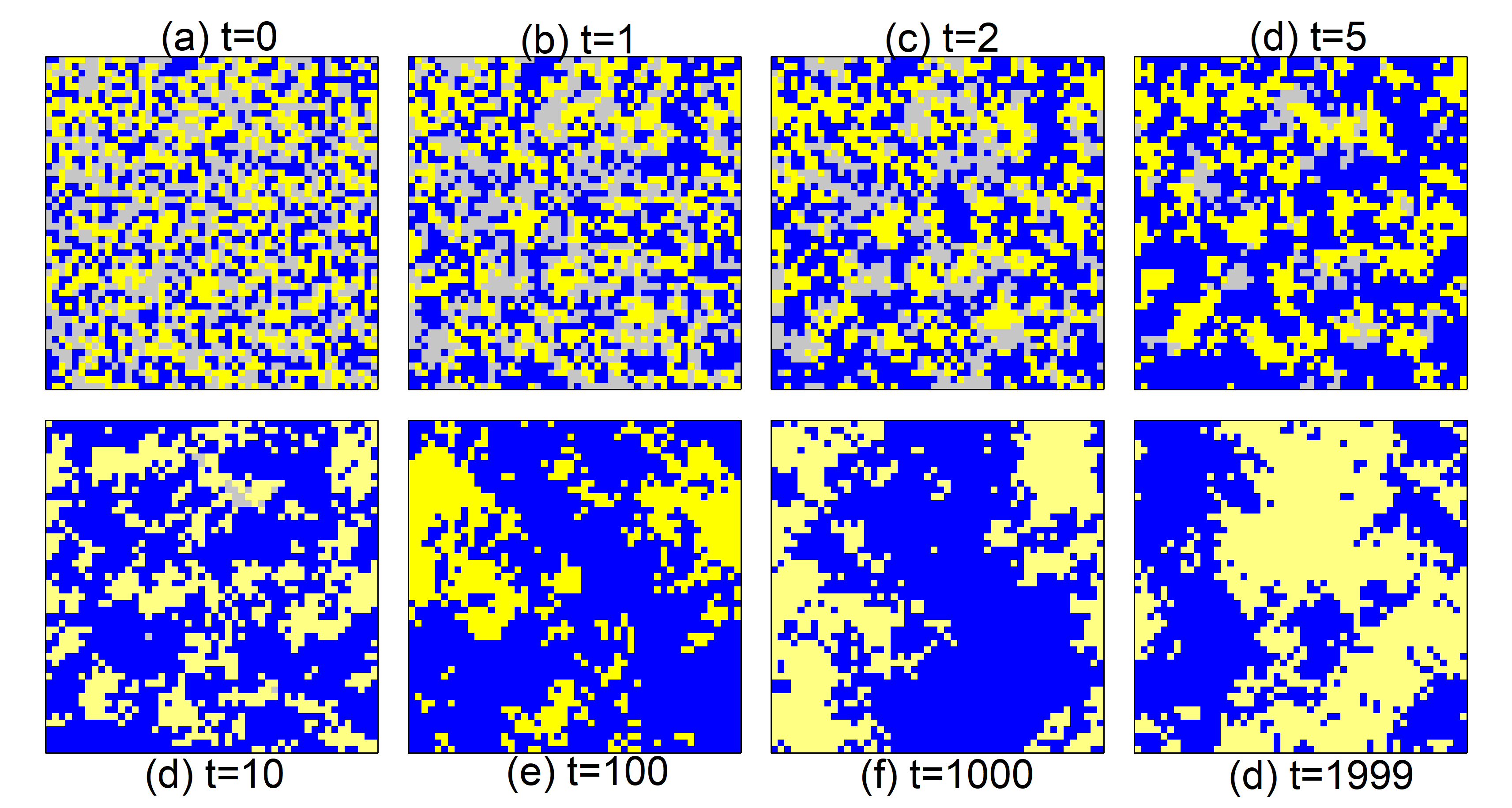}
\caption{Evolution of strategies for the Shunning Rule. The snapshots show the spatial distribution of three different strategists on the square lattice in a representative realization in which DISC strategists follow the Shunning Rule. From panel (a) to panel (h), we record the distribution of ALLC (Yellow dots), ALLD (Gray donts) and DISC (Blue dots), with each panel corresponding to a different time frame $t$ (generation). For this realization, we have taken $H_0=0.2$ and $b=9$. Other parameters are $N=2500$, $c=1$, $w=0.165$, $h=50$, $M=5$, and $K=1$.%Other model parameters are set as the default values in Tab. \ref{tab2}.
}\label{fig:3}
\end{center}
\end{figure}

%\subsection{Model parameters}
In this section, we present and discuss the results of numerical simulations for the agent-based model proposed here. We reduce the payoffs matrix parameters by fixing $c=1$ and
focus on the impact of recipient's benefit $b$ and intolerance threshold %(required minimum reputation) 
$H_0$ on the cooperative behavior under the different second-order assessment rules considered. Based on previous experiments \cite{cuest15_sp}, we fix the additional weight of the last action to $w=0.165$ and the memory length to $M=5$. Each independent realization is run up to %$g=2000$
2000 generations, ensuring that the system can reach a steady state, which is reached typically after 100-1000 generations. Fig. \ref{fig:1}-\ref{fig:5}, which will be discussed in the following subsections, display the results corresponding to $N=2500$ ($L=50$). Additionally, larger lattice sizes (\textit{e.g.}, $N=10^4$) are also tested
and qualitatively equivalent results have been obtained (not shown here for brevity).

\subsection{Level of cooperation and strategies distribution}

\begin{figure}
\begin{center}
\includegraphics[width=1.0\columnwidth]{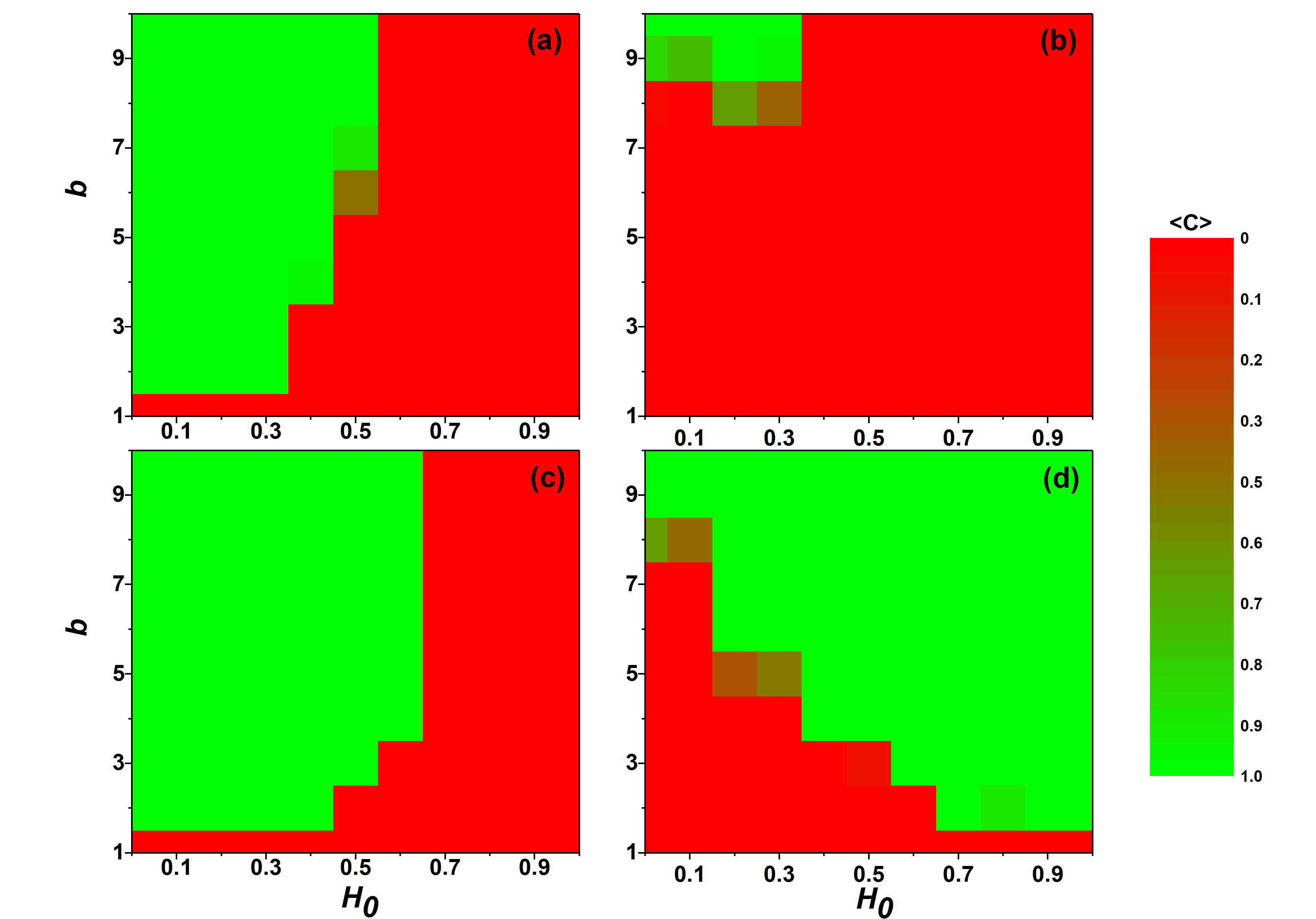}
\caption{Fraction of cooperating actions $\langle C \rangle $ in the stationary state as a function of the benefit $b$ and intolerance $H$, for Shunning (panel a), Stern Judging (b), Image Scoring (c), and Simple Standing (d) assessments rules. All results are averaged over $20$ independent runs. Other parameters are $N=2500$, $c=1$, $w=0.165$, $h=50$, $M=5$, and $K=1$.%Other model parameters are set as the default values in Tab. \ref{tab2}.
}\label{fig:1}
\end{center}
\end{figure}

Fig. \ref{fig:1} displays the stationary fraction of cooperative actions $\langle C \rangle $ as a function of the 
benefit $b$ and intolerance $H_0$, each panel corresponding to each one of the four assessment rules considered: Shunning (panel a), Stern Judging (b), Image Scoring (c) and Simple Standing (d). In general, a very low recipient's benefit ($b\sim1$) does not encourage the donor to donate%, especially for the Simple Standing (panel c) and Stern Judging (d) rules
. For higher values of $b$, the level of cooperation depends on the benefit $b$ and intolerance $H_0$ in different ways for different assessment rules. As shown, a low intolerance $H_0$ promotes  cooperation for Image Scoring and Shunning rules while, conversely, Simple Standing rule behaves better for high values of $H_0$. Finally, Stern Judging rule is the least favorable for cooperation since it only allows cooperative actions for very high benefit $b$ and low intolerance $H_0$. This last result differs from previous studies \cite{ohtsuki2004,ohtsuki2006} where neither memory nor intolerance was considered. %The combination of the memory effect with the second-order evaluation leads to non-trivial impact on the evolution of cooperation.

\begin{figure}
\begin{center}
\includegraphics[width=1.0\columnwidth]{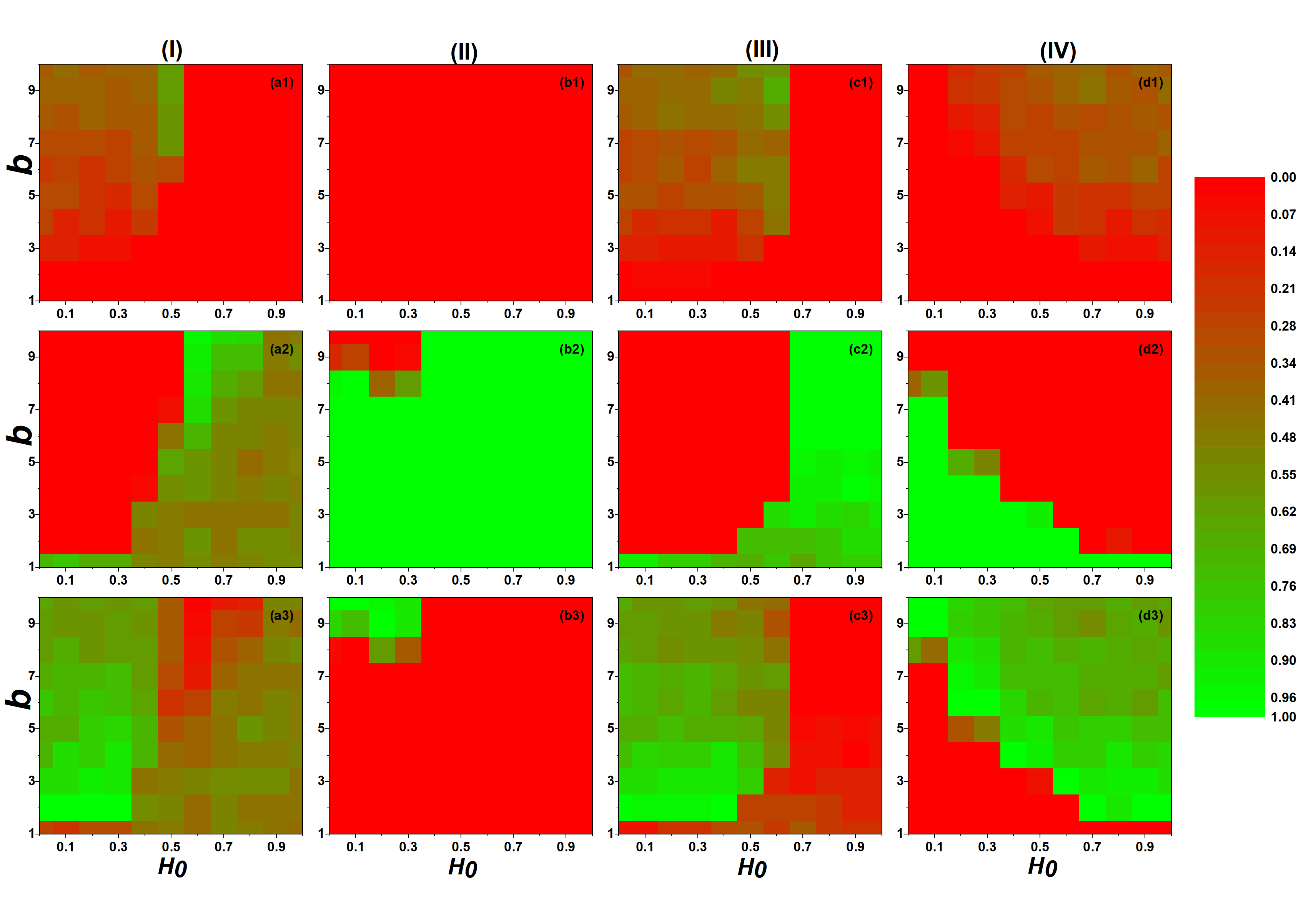}
\caption{Fraction of unconditional cooperators (ALLC), unconditional defectors (ALLD) and discriminators (DISC) at the stationary state as a function of the benefit $b$ and intolerance $H_0$ for four different second-order assessment rules. From left to right, each column corresponds to a different rule: Shunning (column I), Stern Judging (II), Image Scoring (III) and  Simple Standing (IV). Top (\textit{resp.}, center, bottom) row panels display the fraction of ALLC (\textit{resp.}, ALLD, DISC) strategists. All results are averaged over $20$ independent runs. Other parameters are $N=2500$, $c=1$, $w=0.165$, $h=50$, $M=5$, and $K=1$.
See the text for further details. %Other model parameters are set as the default values in Tab. \ref{tab2}.
}\label{fig:2}
\end{center}
\end{figure}

To further study the differences in the cooperation level for the different assessment rules, Figure \ref{fig:2} shows the distribution of the different strategies --ALLC, ALLD, and DISC-- as a function of $b$ and $H_0$. From left to right, each column corresponds to one of the four assessment rules: Shunning (column I), Stern Judging (II), Image Scoring (III) and Simple Standing (IV). 
Additionally, for each column, the fraction of each strategy at the stationary state is shown in different rows: ALLC (panels in top row), ALLD (center), and DISC (bottom).
%Additionally,for each column, the fraction of ALLC, ALLD and DISC strategists at the stationary state is shown in the corresponding panels from top to bottom respectively.
%For each column, the fraction of ALLC, ALLD and DISC strategists at the stationary state is shown in the corresponding panels from top to bottom. As an example, in Column (I), panel (a1), (b1) and (c1) denote the fraction of ALLC, ALLD and DISC under the shunning rule, respectively; and corresponding results for the other $3$ rules are arranged in the right three columns.
Generally speaking, under one specific assessment rule, the level of cooperation is determined by the competition between ALLD, DISC, and ALLC strategists.%; and the cooperation tends to be extinct if ALLD invades the population, otherwise the cooperation will thrive. <- This is not always true, ALLD may coexist with DISC, the later defecting (e.g., Shunning, high H_0)
The coexistence of these three strategies is difficult,
showing (simplex) inner points only for very specific regions of the parameter space.

The distribution of strategies can help explain cooperative behavior for the different rules. Note that the arguments used here, although from a qualitative nature, include more ingredients than those used in the previous mean-field approximation, such as memory and spatial distribution, and the results exhibit some new non-trivial phenomena as follows:

\begin{itemize}
  \item \textbf{Shunning}: Here, DISC players   will only positively evaluate CC actions and therefore do not cooperate against ALLD players. For a low intolerance threshold $H_0$, given an initial homogeneous strategy distribution ($\rho_{c}\sim\rho_{d}\sim\rho_{i}$),
  and a large enough memory (in plots, $M=5$), DISC and ALLC players will have, on average, a fraction $\mu_{S}$ of CC actions in their memory such that $\mu_{S}>H_0$, and therefore will be positively evaluated by DISC players. Thus,
  DISCs will likely cooperate when facing DISC and
  ALLC players. As ALLC strategists will cooperate against any strategist, ALLC and DISC players can group and form cooperative
  clusters for high enough $b$, invading ALLD players
  (who only receive donations from ALLC players). Without ALLD players,
  ALLC and DISC are equivalent strategies and
  will coexist as cooperators.
  On the other hand, for high values of intolerance $H_0$, 
  a small fraction of actions belonging
  to the set $\{DC, CD, DD\}$
  (any strategist is compatible with one or
  more actions in that set and will be likely present in his history at the early stages) %$\rho_{c}\sim\rho_{d} \sim \rho_{i}$)
  is enough to be negatively
  evaluated by other DISC players, 
  that is, any strategist will have, on average, a fraction $\mu_S$ of CC or actions in his memory such that $\mu_S<H_0$, and therefore will be negatively evaluated by DISC players. 
  Thus, DISC strategists will likely not cooperate against any strategist, acting as ALLD players. DISC and ALLD players  (which constitute a majority acting as
  a unique strategy) will invade ALLC and will coexist as defectors.
  %For a low intolerance threshold (\textit{i.e.}, low values of $H_0$), any recipient is more easily evaluated as a good agent and the donor is willing to cooperate with him to enhance the current reputation, and ALLC and DISC strategists seem to be equivalent under this condition, which leads to the situation that ALLC and DISC strategies commonly invade the ALLD strategy; while for a larger $H_0$, the player is often judged to be bad and the defection has an advantage over the cooperation since the donor does not pay anything for this case, which further renders that ALLD and DISC behave equally, and ALLC is commonly invaded by ALLD and DISC in the end so that the cooperation disappears.
  
  \item \textbf{Stern Judging}: In this rule, DISC players will positively evaluate CC and DD actions. For low values of $H_0$, at early stages ($\rho_{c}\sim\rho_{d}\sim\rho_{i}$),
  any strategist will have in his memory, on average, a fraction $\mu_{SJ}$ of 
  actions belonging to the set
  $\{CC,DD\}$ such that $\mu_{SJ}>H_0$, and therefore will be positively evaluated by DISC players. Thus,
  DISC players will likely cooperate when facing any strategist, behaving as ALLC players. 
  Unlike the previous Shunning case, where ALLD players obtained benefit only from ALLC, now they get donations both from ALLC and DISC players, thus having a higher relative payoff and resulting in an invasion of ALLD over the rest of strategies. Only for very high values of benefit (in Fig. \ref{fig:1}-\ref{fig:2}, $b\gtrsim9$) DISC players can resist invasion by ALLD; actually, this is the only region of parameters that allows cooperation. On the other hand, for high values of intolerance $H_0$, and at early stages, %($\rho_{c}\sim\rho_{d}\sim\rho_{i}$), 
  any strategist will have, on average, a fraction $\mu_{SJ}$ of 
  actions belonging to the set
  $\{CC,DD\}$ in his memory such that $\mu_{SJ}<H_0$, and therefore will be negatively evaluated by DISC players. However, although DISC players tend to defect against any strategist, they have a non-zero probability of cooperating when facing any player (either ALLC, ALLD, or DISC), resulting in a lower accumulated payoff than that of ALLD players. Therefore, DISC players will get the higher accumulated payoff, which drives to an invasion of ALLD strategy over ALLC and DISC.

  %The cooperation can be fostered only provided that $b$ is high enough and $H_0$ is small enough. Similarly, for a smaller $H_0$, most players can maintain the good reputation through the cooperating action, which leads to the higher benefit for ALLD-players on average, who will take over the population except that $b$ is high enough (e.g., $b=9$ or $b=10$); For a larger $H_0$, the player is difficult to be assessed as good, while the actor is encouraged to defect the bad recipient to keep the good status under this rule, meanwhile ALLC strategy is always exploited by the ALLD one paying nothing, which eventually dominates the whole population.
  
  \item \textbf{Image Scoring}: Here, DISC players
  will positively evaluate CC and CD actions and therefore will cooperate against ALLC but not against ALLD players. For low values of intolerance $H_0$, at early stages, %($\rho_{c}\sim\rho_{d}$\sim$\rho_{i}$),
  DISC players will have, on average, a fraction $\mu_{IS}$ of actions
  cooperative actions (CC or CD) in their memory such that $\mu_{IS}>H_0$, and therefore will be positively evaluated by other DISC players. Thus, DISCs will likely cooperate when facing DISC and ALLC players. This is the same situation than that corresponding to the previous Shuning - low $H_0$ case:
  ALLC and DISC players can form cooperative
  clusters for high enough $b$, invading ALLD strategy.
  Without ALLD players, ALLC and DISC will coexist as cooperators. On the other hand, for high values of intolerance $H_0$, at early stages,
  DISC players will have, on average, a fraction $1-\mu_{IS}$ of
  non-cooperative actions in their memory such that $\mu_{IS}<H_0$, and therefore will likely be negatively evaluated by other DISC players. Each strategist will cooperate against a different set of strategies: ALLC against any strategist, DISC against ALLC ones, and ALLD against no strategist. This results in a three-strategies scenario where the higher payoff corresponds to ALLD players, who will invade ALLC and ALLD.
  
  %The cooperation displays the same behavior as the Shunning rule, but the cooperation region is further widened. As the goodness threshold $H_0$ becomes higher and the player is more easily evaluated as bad. Although the cooperating action can help to build the good reputation for a donor at this time, most of players may be still thought to be a bad one as a result of weighted score and larger $H_0$. Then, this rule attracts the DISC player to give up the cooperation in the future. Finally, ALLD strategy acquires the evolutionary advantage and even renders the ALLC and DISC strategies to be completely invaded by ALLD one.
  
  \item \textbf{Simple Standing}: This rule is the most tolerant: the only negatively evaluated action is defecting against a cooperator. Counterintuitively, while high values of intolerance $H_0$ promote cooperation, low values drive to non-cooperative states.
  Here, DISC players will positively evaluate CC, CD and DD actions. As the available actions for ALLC players are $\{CC, CD\}$, and those for DISC ones are $\{CC, CD, DD\}$, DISC strategists will always cooperate against any ALLC or DISC player.
  For low values of $H_0$, at early stages,
  ALLD players will have, on average, a fraction $\mu_{SS}$ of DD actions in their memory such that $\mu_{SS}>H_0$, and therefore will be positively evaluated by DISC strategists. 
  This is the same situation than that corresponding to the previous Stern Judging - low $H_0$ case: DISC players will behave as ALLC players. ALLD players will obtain the higher accumulated payoff, resulting in an invasion of ALLD over the rest of strategies. Only for very high values of $b$, DISC players can resist invasion by ALLD. As $H_0$ increases, $\mu_{SS}-H_0$ decreases, moving towards the following scenario: for high values of intolerance $H_0$, 
  ALLD players will have, on average, a fraction $\mu_{SS}$ of DD actions at the early stages such that $\mu_{SS}<H_0$, and therefore
  will be negatively evaluated by DISC players. 
  Thus, DISC players will cooperate when facing DISC and
  ALLC, but very unlikely when facing ALLD players. As ALLC strategists will cooperate against any player, ALLC and DISC players can form cooperative clusters for high enough $b$, invading ALLD strategists
  who only receive donations from ALLC. In the absence of ALLD players, ALLC and DISC strategists will behave alike and will coexist.
  
  %The rule is the most tolerant one and attracts the DISC ones to perform the donation. The cooperation can be greatly facilitated within most of the parameter space, and the cooperation will be inhibited only within the left-bottom region, where $H_0$ is very small and $b$ is also lower. At this case, ALLC and DISC players are almost equivalent and always cooperating so that the ALLD recipient can obtain the donation benefit and accumulate the more advantage, and then ALLD strategy invades ALLC and DISC ones during the strategy update. However, for other regions, the Simple Standing fosters the evolution of cooperation since ALLC and DISC are equal under this situation and then create the compact cooperative clusters to commonly resist the invasion of ALLD strategy, or even punish the ALLD players. Around the diagonal regions, DISC strategy even dominates the whole population and maintains the high cooperation.
\end{itemize}

\subsection{Spatial patterns}
\begin{figure}
\begin{center}
\includegraphics[width=1.0\columnwidth]{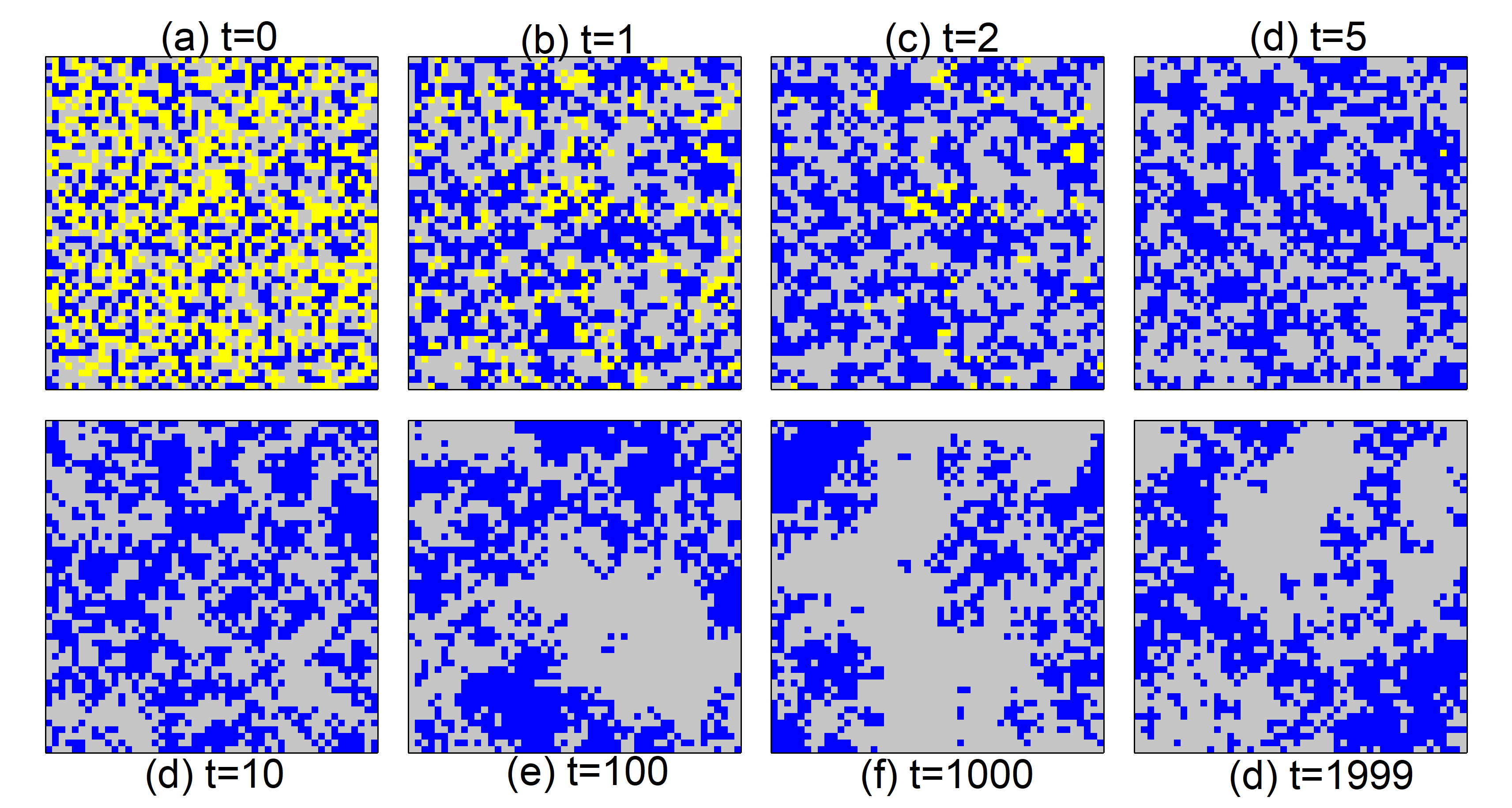}
\caption{Evolution of strategies under the Shunning Rule. From panel (a) to panel (h), we record the distribution of ALLC (Yellow dots), ALLD (Gray) and DISC (Blue). Each panel corresponds to a different time frame $t$ (generation). In this characteristic simulation, $H_0=0.9$ and $b=4$. Other parameters are $N=2500$, $c=1$, $w=0.165$, $h=50$, $M=5$, and $K=1$.%Other model parameters are set as the default values in Tab. \ref{tab2}.
}\label{fig:4}
\end{center}
\end{figure}

In this subsection, we analyze the strategies evolution through characteristic snapshots. Taking the Shunning Rule as an example, Fig. \ref{fig:3} and Fig. \ref{fig:4} display the strategies distribution on the square lattice at different generations ($t$) to scrutinize the evolutionary process, for a low intolerance ($H_0=0.2$) in Fig. \ref{fig:3} and large ($H_0=0.9$) in Fig. \ref{fig:4}. As mentioned above, for low values of $H_0$, ALLC and DISC players are more easily to be positively evaluated: DISC strategists will cooperate when facing DISC and ALLC players, allowing the formation of cooperative clusters. In Fig. \ref{fig:3}, it is shown that ALLD strategists (gray dots) are gradually invaded by ALLC and DISC ones, and they disappear after around $100$ generations. Even if we reduce the benefit of the recipient (say, $b=2$ or $3$), similar patterns are still observed (although they are not shown here for the sake of shortness). However, when $H_0$ is large enough (\textit{e.g.}, $H_0\gtrsim0.5$), ALLC and DISC players are often negatively evaluated, which leads DISCs to act as ALLD strategists. %, resulting in an invasion over ALLC strategy. 
Thus, for largue $H_0$, ALLD and DISC strategies are equivalent and invade ALLC. This behavior is shown in Fig. \ref{fig:4}, where gray (ALLD) and blue (DISC) dots dominate the whole population just after $10$ generations.

\begin{figure}
\begin{center}
\includegraphics[width=1.0\columnwidth]{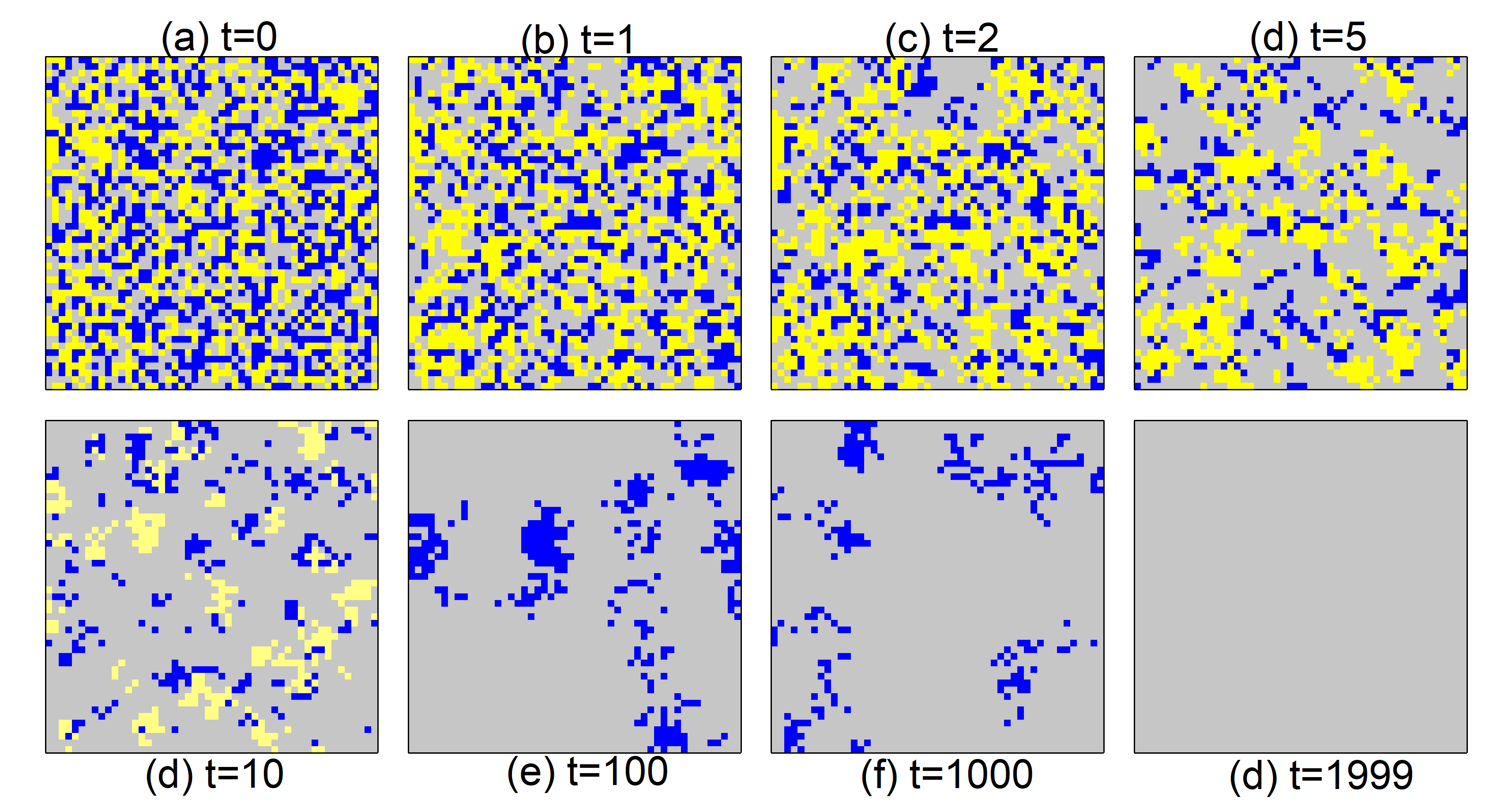}
\caption{Time evolution of three different strategies under the Image Scoring Rule. From panel (a) to panel (h), we record the distribution of ALLC (Yellow dots), ALLD (Gray) and DISC (Blue dots), with each panel corresponding to a different time frame. In this characteristic realizations, we have taken $H_0=0.9$ and $b=4$. Other parameters are $N=2500$, $c=1$, $w=0.165$, $h=50$, $M=5$, and $K=1$.%Other model parameters are set as the default values in Tab. \ref{tab2}.
}\label{fig:5}
\end{center}
\end{figure}

Similarly, the Image Scoring Rule also creates the coexistence between ALLC and DISC ones for low values of the intolerance $H_0$, and even this case appears for larger $H_0$ when compared to the Shunning Rule (the corresponding characteristic snapshots are not shown here for the sake of brevity). Conversely, for high intolerance $H_0$, as mentioned above, DISC players cooperate when facing ALLC but not DISC ones. Thus, ALLD players obtain the higher benefit and dominate the whole population. This evolutionary process is illustrated in Fig. \ref{fig:5}, where the simulation setup is identical to that of Fig. \ref{fig:4} except that here applies the Image Scoring Rule. 
The same approach can be used to characterize the competition among three strategies under Stern Judging and Simple Standing rules (not shown here for conciseness).

\section{Discussion and conclusions}
\label{Section:DicussionAndConclusions}
In this paper, we combine four typical second-order assessment rules with the memory effect to explore the evolution of cooperation in the spatial donation game. To this end, the reputation evaluation takes into account the last $M$ actions of the agents. We discuss the impact of four assessment rules --namely Shunning, Stern Judging, Image Scoring, and Simple Standing-- on the level of cooperation among the population. It is found that the assessment rule plays a non-trivial role in the evolution of cooperation.

In our model, the interplay between any pair of players can be characterized as a Donation Game, where a player is chosen as a donor and the other one as a recipient. If the donor contributes by paying a cost $c$, the recipient will get a benefit $b>c$; otherwise, both will get nothing. We implement two dynamics, one in which strategies do not take into account neighbors' payoffs but their reputation,
modulated by an intolerance parameter,
and another evolutionary dynamic that takes place at a larger time scale.

%In order to help a donor to make the decision within one period, we need to record the recent $M=5$ actions for each individual as the basis of reputation evaluation. Then, we can calculate the weighted score according to the last action's score and the average score of all $M$ actions under a specific assessment rule. After that, we can compare this score with a model parameter $H_0$, which is termed as the goodness threshold, and then judge whether he is good or bad. If the recipient is evaluated as good, the donor will make the donation, or else he will not contribute.

We have studied the model through a mean-field approximation, finding that the role of intolerance varies according to the assessment rule: while under Shunning, Stern Judging and Image Scoring rules intolerance hinders cooperation, it counterintuitively promotes it under Simple Standing rule. Moreover, it is shown that Stern Judging rule, despite being a positive rule (positively evaluates more actions than the Shunning Rule), is the one that shows, by far, the lowest values of cooperation.
We have performed extensive simulations that confirm these findings.

%In addition, it is also strongly indicated that the assessment rule can play an important role in the evolution of cooperation and create the non-trivial influence. For the shunning and Image Scoring rules, the cooperation can reach a very high level when $H_0$ is smaller (e.g. $H_0\le 0.5$ for the shunning rule and $H_0\le 0.7$ for the image one); As $H_0$ becomes larger, the cooperation is abruptly decreased and even the full defection is reached. Regarding the Simple Standing rule, except that $H_0$ is small and $b$ is not too high, the cooperation can be greatly fostered under most ranges of $H_0-b$ parameter space, and the current results also conform to the intuition since the Simple Standing rule is the most tolerant one.

%However, the Stern Judging rule leads to the full defection for most combinations of $H_0$ and $b$, and supports the cooperating behavior only when $H_0$ is very small ($H_0\le 0.2$) and the benefit parameter $b$ is very high (e.g., $b=9$ or $b=10$) although the Stern Judging rule is evolutionarily stable, which is also consistent with those of Sasaki \textit{et al.} \cite{sasaki17_games}, and they pointed out that the cooperation will not be promoted under the direct observation and private assessment against the individual reputation. In our model, each donor will directly judge whether the recipient is good or not by comparing the weighted score and the goodness threshold $H_0$, which is rightly akin to the direct observation and private assessment in Ref. \cite{sasaki17_games}.

Furthermore, there are several other parameters (\textit{e.g.}, noise parameter $K$ and memory length $M$) that deserve consideration in future research. As $K$ increases, the strategy adoption uncertainty is also increased, but the level of cooperation can still be qualitatively kept unchanged in the current setup. With regard to the impact of memory length or weight, we only adopt the parameter values ($M=5$ and $w=0.165$) in Ref.\cite{cuest15_sp}, but it may deserve further discussion in the future. Meanwhile, observation or reputation evaluation errors may take place during the decision of donation, which is also worth being further investigated in future studies. Another potential direction could be conducted to explore the impact of second-order assessment rules in heterogeneous networks, such as small-world \cite{watts98_nature} and scale-free \cite{barabasi99_sci} networks.

Taking together, based on previous experimental findings on human behavior, we present a novel second-order evaluation model with memory effect to investigate the evolution of cooperation in the spatial donation game. These results may help to understand the cooperative behavior under the indirect reciprocity and reputation mechanisms.

\begin{acknowledgments}
CYX acknowledges the financial support of the National Natural Science Foundation of China (NSFC) under Grant No. 61773286, and the support of China Scholarship Council under Grant No. 201808120001.  CGL and YM acknowledge partial support from Project UZ-I-2015/022/PIP, the Government of Arag\'on, Spain, and FEDER funds, through grant E36-17R to FENOL, and from MINECO and FEDER funds (grant FIS2017-87519-P). The funders had no role in study design, data collection, and analysis, decision to publish, or preparation of the manuscript.
\end{acknowledgments}

\nocite{*}
%\bibliography{aipsamp}% Produces the bibliography via BibTeX.

\begin{thebibliography}{1}%
\makeatletter
\providecommand \@ifxundefined [1]{%
 \@ifx{#1\undefined}
}%
\providecommand \@ifnum [1]{%
 \ifnum #1\expandafter \@firstoftwo
 \else \expandafter \@secondoftwo
 \fi
}%
\providecommand \@ifx [1]{%
 \ifx #1\expandafter \@firstoftwo
 \else \expandafter \@secondoftwo
 \fi
}%
\providecommand \natexlab [1]{#1}%
\providecommand \enquote  [1]{``#1''}%
\providecommand \bibnamefont  [1]{#1}%
\providecommand \bibfnamefont [1]{#1}%
\providecommand \citenamefont [1]{#1}%
\providecommand \href@noop [0]{\@secondoftwo}%
\providecommand \href [0]{\begingroup \@sanitize@url \@href}%
\providecommand \@href[1]{\@@startlink{#1}\@@href}%
\providecommand \@@href[1]{\endgroup#1\@@endlink}%
\providecommand \@sanitize@url [0]{\catcode `\\12\catcode `\$12\catcode
  `\&12\catcode `\#12\catcode `\^12\catcode `\_12\catcode `\%12\relax}%
\providecommand \@@startlink[1]{}%
\providecommand \@@endlink[0]{}%
\providecommand \url  [0]{\begingroup\@sanitize@url \@url }%
\providecommand \@url [1]{\endgroup\@href {#1}{\urlprefix }}%
\providecommand \urlprefix  [0]{URL }%
\providecommand \Eprint [0]{\href }%
\providecommand \doibase [0]{http://dx.doi.org/}%
\providecommand \selectlanguage [0]{\@gobble}%
\providecommand \bibinfo  [0]{\@secondoftwo}%
\providecommand \bibfield  [0]{\@secondoftwo}%
\providecommand \translation [1]{[#1]}%
\providecommand \BibitemOpen [0]{}%
\providecommand \bibitemStop [0]{}%
\providecommand \bibitemNoStop [0]{.\EOS\space}%
\providecommand \EOS [0]{\spacefactor3000\relax}%
\providecommand \BibitemShut  [1]{\csname bibitem#1\endcsname}%
\let\auto@bib@innerbib\@empty
%</preamble>
\bibitem [{Note1()}]{Note1}%
  \BibitemOpen
  \bibinfo {note} {Cuesta \protect \textit {et al.} \cite {cuest15_sp} showed
  through lab-based human experiments that people measures reputation based on
  the weighted average of the fraction of cooperative actions ($\protect
  \mathaccentV {bar}016{C}$) and the last action performed ($C_{last}$), in
  which the coupling relationship can be linearly characterized as
  $w*C_{last}+(1-w)*\protect \mathaccentV {bar}016{C}$ and $w$ is often fitted
  to be $0.165$ from the experimental data.}\BibitemShut {Stop}%
\end{thebibliography}%


\begin{thebibliography}{aipsamp}

\bibitem{pennisi2005} E. Pennisi, "How did cooperative behavior evolve?", Science \textbf{309}, 93 (2005).
\bibitem{axelrod06} R. Axelrod, "The evolution of cooperation", Basic Books, New York, NY (2006).
\bibitem{sigmund00} K. Sigmund, "The calculus of selfishness", Princeton University Press, Princeton, NJ (2010).
\bibitem{nowak_06book} M.A. Nowak, "Evolutionary dynamics: Exploring the equations of life", Harvard Universtiy Press, Cambrige, MA (2006).

\bibitem{szab1997} G. Szab\'o, and G. F\'ath, "Evolutionary games on graphs", Phys. Rep. \textbf{446}, 97-216 (2007).
\bibitem{perc2010} M. Perc, and A. Szolnoki, "Coevolutionary games--A mini review", BioSystems \textbf{99}, 109-125 (2010).
\bibitem{perc2013} M. Perc, J. G\'omez-Garde\~nes, A. Szolnoki, L.M. Flor\'ia, and Y. Moreno, "Evolutionary dynamics of group interactions on structured populations: a review", J. R. Soc. Interface \textbf{10}, 20120997 (2013).
\bibitem{zhen2015_epjb} Z. Wang, L. Wang, A. Szolnoki, and M. Perc, "Evolutionary games on multilayer networks: a colloquium", The European physical journal B \textbf{88}, 124 (2015).

\bibitem{nowak2006} M.A. Nowak, "Five rules for the evolution of cooperation", Science \textbf{314}, 1560-1563 (2006).
\bibitem{Hamilton64} W.D. Hamilton, "The genetical evolution of social behaviour", J. Theor. Biol. \textbf{7}, 1--16 (1964).
\bibitem{Trivers71} R.L. Trivers, "The evolution of reciprocal altruism", Quat. Rev. Biol. \textbf{46}, 35 (1971).
\bibitem{nowak1998a} M.A. Nowak, and K. Sigmund, "Evolution of indirect reciprocity by Image Scoring", Nature \textbf{393}, 573-577 (1998).
\bibitem{Traulsen06} A. Traulsen, and M.A. Nowak. "Evolution of cooperation by multilevel selection", Proc. Natl. Acad. Sci. (USA) \textbf{103},
10952-10955 (2006).

\bibitem{nowak1992} M.A. Nowak, and R.M. May, "Evolutionary Games and Spatial Chaos", Nature \textbf{359}, 826--829 (1992).
\bibitem{santos2005} F.C. Santos, and J.M. Pacheco. "Scale-free networks provide a unifying framework for the emergence of cooperation", Phys. Rev. Lett. \textbf{95}, 098104 (2005).
\bibitem{yamir2007_prl} J. G\'omez-Garde\~nes, M. Campillo, L.M. Flor\'ia, and Y. Moreno, "Dynamical organization of cooperation in complex topologies", Phys. Rev. Lett. \textbf{98}, 108103 (2007).
\bibitem{yamir2012_pnas} C. Gracia-L\'azaro, A. Ferrer, G. Ruiz, A. Taranc\'on, J.A. Cuesta, A S\'anchez, and Y. Moreno, "Heterogeneous networks do not promote cooperation when humans play a Prisoner's Dilemma", Proc. Natl. Acad. Sci. (USA) \textbf{109}, 12922-12926 (2012).

\bibitem{nowak05_review} M.A. Nowak, and K. Sigmund, "Evolution of indirect reciprocity", Science \textbf{437}, 1291-1298 (2005).
\bibitem{marsh18_review} C. Marsh, "Indirect reciprocity and repuation management: Interdisciplinary findings from evolutionary biology and economics", Pub. Relations Rev. \textbf{44}, 463-470 (2018).

\bibitem{ohtsuki2004} H. Ohtsuki, and Y. Iwasa, "How shold we define goodness?--reputation dynamics in indirect reciprocity", J. Theor. Biol. \textbf{231}, 107-120 (2004).
\bibitem{wang2012_plos} Z. Wang, Z.Y. Yin, L. Wang, and C.Y. Xia, "Inferring reputation promotes the evolution of cooperation in spatial social dilemma games", PLoS ONE \textbf{7}, e40218 (2012).
\bibitem{xia2016_pla} M.H. Chen, Z. Wang, S.W. Sun, J. Wang, and C.Y. Xia, "Evolution of cooperation in the spatial public goods game with adaptive reputation assortment", Phys. Lett. A \textbf{380}, 40-47 (2016).
    
\bibitem{uchida2010} S. Uchida, and K. Sigmund, "The competition of assessment rules for indirect reciprocity", J. Theor. Biol. \textbf{263}, 13-19 (2010).

\bibitem{reward1} R. Jim\'enez, H. Lugo, J.A. Cuesta, and A.S\'anchez, "Emergence and resilience of cooperation in the spatial prisoner's dilemma via a reward mechanism", J. Thero. Bio. \textbf{250}, 475-483 (2008).
\bibitem{reward2} A. Szolnoki, and M. Perc, "Antisocial pool rewarding does not deter public cooperation", Proc. Roy. Soc. B \textbf{282}, 20151975 (2015).

\bibitem{punish1} E. Fehr, and S. G\"{a}chter, "Altruistic punishment in humans", Nature \textbf{415}, 137-140 (2002).
\bibitem{punish2} T. Sasaki, I. Okada, S. Uchida, and X. Chen, "Commitment to cooperation and peer punishment: Its evolution", Games, \textbf{6}, 574-587 (2015).
\bibitem{punish3} Z. Wang, C.Y. Xia, S.Meloni, C.S. Zhou, and Y. Moreno, "Impact of social punishment on cooperative behavior in complex networks", Sci. Rep. \textbf{3}, 3055 (2013).

\bibitem{nowak1998b} M.A. Nowak, and K. Sigmund, "The dynamnics of indirect reciprocity", J. Theor. Biol. \textbf{194}, 561-574 (1998).

\bibitem{leimar2001} O. Leimar, and P. Hammerstein, "Evolution of cooperation through indirect reciprocity", Proc. R. Soc. Lond. B \textbf{268}, 745-753 (2001).
\bibitem{sugden1986} R. Sugden, "The economics of rights, cooperation and welfare", Oxford, UK: Basil Blackwell (1986).
\bibitem{body2003} K. Panchanathan, and R. Boyd, "A tale of two defectors: the importance of standing for evolution of indirect reciprocity", J. Theor. Biol. \textbf{224}, 115-126 (2003).
\bibitem{ohtsuki2006} H. Ohtsuki, and Y. Iwasa, "The leading eight: social norms that can maintain cooperation by indirect reciprocity", J. Theor. Biol. \textbf{239}, 435-444 (2006).
\bibitem{milinski2001} M. Milinski, D. Semmann, T.C.M. Bakker, and H.J. Krambeck H, "Cooperation through indirect reciprocity: Image Scoring or standing strategy?", Proc. R. Soc. Lond. B \textbf{268}, 2495-250 (2001).
\bibitem{bolton2004} G.E. Bolton, E. Katok, A. Ockenfels, "Cooperation among strangers with limited information about reputation", J. Public Econ. \textbf{89}, 1457-1468 (2005).

\bibitem{uchida2010_pre} S. Uchida, "Effect of private information on indirect reciprocity", Phys. Rev. E \textbf{82}, 036111 (2010).
\bibitem{Nakamaru2004} M. Seki, and M. Nakamaru, "Model for gossip-mediated evolution of altruism with various types of false information by speakers and assessment by listerners", J. Theor. Biol. \textbf{470}, 90-105 (2016).

\bibitem{sasaki17_games} T. Sasaki, H. Yamamoto, I. Okdada, and S. Uchida, "The evolution of reputation-based cooperation in regular networks", Games \textbf{8}, 8 (2017).

\bibitem{memory_pre} W.X. Wang, J. Ren, G.R. Chen, and B.H. Wang, "Memory-based snowdrift game on networks", Phys. Rev. E \textbf{74}, 056113 (2006).

\bibitem{cuest15_sp} J.A. Cuesta, C. Gracia-L\'azaro, A. Ferrer, Y. Moreno, and A. S\'anchez, "Reputation drives cooperative behavior and network formation in human groups", Sci. Rep. \textbf{5}, 7843 (2015).


\bibitem{fermi98} G. Szab\'o, and C. T\"oke, "Evolutionary prisoner's dilemma game on a square lattice", Phys. Rev. E \textbf{58}, 69-73 (1998).
\bibitem{diversity2} M. Perc, "Does strong heterogeneity promote cooperation by group interactions?" New J. Phys. \textbf{13}, 123027 (2011).
\bibitem{diversity3} A. Szolnoki, M. Perc, and G. Szab\'o, "Topology-independent impact of noise on cooperation in spatial public goods games", Phys. Rev. E \textbf{80}, 056109 (2009).

\bibitem{Moran1962} P.A.P. Moran, "The statistical processes of evolutionary theory", Oxford, UK: Clarendon Press (1962).

\bibitem{watts98_nature} D. Watts, and S. Strogatz, "Collective dynamics of small-world networks", Nature \textbf{393} 440--442 (1998).
\bibitem{barabasi99_sci} A. Barabasi, and R. Albert, "Emergence of scaling in random networks" Science \textbf{286}, 509--512 (1999).

\bibitem{gracia12_SciRep} C. Gracia-L\'azaro, J.A. Cuesta, A. S\'anchez, and Y. Moreno, "Human behavior in Prisoner's Dilemma experiments suppresses network reciprocity." Sci. Rep. \textbf{2}, 325 (2012).

\end{thebibliography}

\end{document}